\documentclass[amsmath, amssymb, preprintnumbers, showpacs, showkeys, aps,prb,superscriptaddress,twocolumn]{revtex4-1}
\usepackage[utf8]{inputenc}
\usepackage{listings}
\usepackage{footmisc}
\usepackage{enumerate}
\usepackage{latexsym}
\usepackage{braket}
\usepackage{graphicx}
\usepackage[caption=false]{subfig}
\usepackage[colorlinks=True,linkcolor=red,citecolor=blue,urlcolor=blue]{hyperref}
\usepackage{blkarray}
\usepackage{array}
\usepackage{color}
\usepackage[normalem]{ulem}

\newcommand{\ms}{\mathsf}
\newcommand{\bs}{\boldsymbol}

\newcommand{\ua}{\uparrow}
\newcommand{\da}{\downarrow}
\newcommand{\dg}{\dagger}

\definecolor{bleu}{rgb}{0.0, 0.5, 0.69}

\date{\today}

\begin{document}


\title{Towards a Topological Quantum Chemistry description of correlated systems: the case of the Hubbard diamond chain}

\author{Mikel Iraola}
\email{mikel.i.iraola@gmail.com}
\affiliation{Donostia International Physics Center, 20018 Donostia-San Sebastian, Spain}
\affiliation{Department of Condensed Matter Physics, University of the Basque Country UPV/EHU, Apartado 644,
48080 Bilbao, Spain}
\author{Niclas Heinsdorf}
\affiliation{Institute of Theoretical Physics, Goethe University Frankfurt,
Max-von-Laue-Straße 1, 60438 Frankfurt am Main, Germany}
\author{Apoorv Tiwari}
\affiliation{Condensed Matter Theory Group, Paul Scherrer Institute, CH-5232 Villigen PSI, Switzerland}
\affiliation{Department of Physics, University of Zurich, Winterthurerstrasse 190, CH-8057 Zurich, Switzerland}
\author{Dominik Lessnich}
\affiliation{Institute of Theoretical Physics, Goethe University Frankfurt,
Max-von-Laue-Straße 1, 60438 Frankfurt am Main, Germany}
\author{Thomas Mertz}
\affiliation{Institute of Theoretical Physics, Goethe University Frankfurt,
Max-von-Laue-Straße 1, 60438 Frankfurt am Main, Germany}
\author{Francesco Ferrari}
\affiliation{Institute of Theoretical Physics, Goethe University Frankfurt,
Max-von-Laue-Straße 1, 60438 Frankfurt am Main, Germany}
\author{Mark H. Fischer}
\affiliation{Department of Physics, University of Zurich, Winterthurerstrasse 190, CH-8057 Zurich, Switzerland}
\author{Stephen M. Winter}
\affiliation{Institute of Theoretical Physics, Goethe University Frankfurt,
Max-von-Laue-Straße 1, 60438 Frankfurt am Main, Germany}
\affiliation{Department of Physics,
Wake Forest University, 1834 Wake Forest Road Winston-Salem, NC 27109-7507, USA}
\author{Frank Pollmann}
\affiliation{Department of Physics and Institute for Advanced Study,
Technical University of Munich, 85748 Garching, Germany}
\author{Titus Neupert}
\affiliation{Department of Physics, University of Zurich, Winterthurerstrasse 190, CH-8057 Zurich, Switzerland}
\author{Roser Valent\'\i}
\email{valenti@itp.uni-frankfurt.de}
\affiliation{Institute of Theoretical Physics, Goethe University Frankfurt,
Max-von-Laue-Straße 1, 60438 Frankfurt am Main, Germany}
\author{Maia G. Vergniory}
\email{maiagvergniory@dipc.org}
\affiliation{Donostia International Physics Center, 20018 Donostia-San Sebastian, Spain}
\affiliation{IKERBASQUE, Basque Foundation for Science, Maria Diaz de Haro 3, 48013 Bilbao, Spain}

\begin{abstract}

The recently introduced topological quantum chemistry (TQC) framework has
	provided a description of  universal topological properties of all possible band insulators in all space groups based on crystalline unitary symmetries and time reversal. While this formalism filled the gap between the mathematical classification and the practical diagnosis of
	topological materials, an obvious limitation is that it only applies to
	weakly interacting systems—which can be described within band theory.
	It is an open question to which extent this formalism can be
	generalized to correlated systems that can exhibit symmetry protected
	topological phases which are not adiabatically connected to any band
	insulator. In this work we address the many facettes of this question by considering the specific example of a Hubbard diamond chain. This model features a Mott insulator, a trivial insulating phase	and an obstructed atomic limit phase. Here we discuss the nature of the Mott insulator and determine the phase
	diagram and topology of the interacting model with infinite density matrix renormalization group calculations, variational Monte Carlo	simulations and with  many-body topological invariants. We then proceed by considering a generalization
	of the TQC formalism to Green's functions
	combined with the concept of topological   
        Hamiltonian to identify the topological nature 
	of the phases, using
	cluster perturbation theory to calculate the Green's functions. 
	The results are benchmarked with the above determined phase diagram and we discuss the applicability and limitations of the approach	and its possible extensions.

\end{abstract}

\maketitle

\section{Introduction}

Topology is one of the central concepts in the modern understanding of electronic quantum matter\cite{CharlieZahidReview,PhysRevLett.95.146802,Bernevig1757}. One of its first incarnations, the quantum Hall effect\cite{RevModPhys.58.519}, demonstrates impressively that topological phenomena can be intimately linked, yet vastly different, depending on whether or not electron-electron interactions are required for them to exist: The integer quantum Hall effect can be understood from non-interacting electrons, while its fractional counter-part\cite{PhysRevLett.48.1559,RevModPhys.71.891} is intrinsically interacting. 

This division between the interacting and non-interacting view on topology got reinforced as the characterization and classification of topological phases advanced substantially over the past years. In particular, the role of symmetries, both of spatial (crystalline) and of global type (e.g., time reversal) has been explored in great depth in both domains\cite{Bradlyn_TQC,ZhidaSemimetals,AshvinIndicators,ChenTCI,Rachel_2018}. 

For interacting phases the notion of symmetry protected topology (SPT)\cite{PhysRevB.80.155131,PhysRevB.81.064439,PhysRevB.83.075103,PhysRevB.83.075102,PhysRevB.83.035107,shentil-rev} 
is central. 
These phases are often defined from a quantum circuit perspective: Two SPT quantum states are distinct if there is no finite depth circuit of local and symmetry-respecting unitary operators that transform into one another. A trivial SPT state is a direct product of local basis states. Spatial, i.e., non-local symmetries have been incorporated in this framework as well. The classification is then derived from that of subdimensional systems that are left invariant under the symmetry, such as a point under inversion symmetry or a plane in the case of three-dimensional mirror symmetry\cite{PhysRevX.7.011020}.  With this, the understanding of SPT phases arises from a local perspective on the physical system, furnished by the degrees of freedom that form the trivial product state and are acted upon by sequences of local unitary operations. 

While the notion of SPTs applies in principle also in the absence of interactions, a much more efficient way of detecting and classifying topology of non-interacting electron systems arises from band theory. Two band insulators are topologically distinct if they cannot be smoothly deformed into one another without breaking a set of protecting symmetries or closing the band gap\cite{PhysRevLett.95.146802,PhysRevLett.98.106803}. 
A wealth of topological invariants can be defined from the fiber bundles of Bloch functions over the Brillouin zone (BZ) to detect topological distinctions\cite{PhysRevB.55.1142,RevModPhys.88.035005}. Examples include Chern numbers for quantum Hall states, winding numbers, and Pfaffian invariants for topological insulators. 
When spatial symmetries are also taken into account, this classification is refined and topological invariants can be formulated from the irreducible symmetry group representations of the Bloch wave functions of occupied bands\cite{PhysRevLett.106.106802,PhysRevB.85.115415}. 

Recently, topological quantum chemistry (TQC) emerged as a new perspective on the topology of non-interacting electronic states\cite{Bradlyn_TQC,vergnio}. Different from topological band theory, it starts from a real-space description with the realization that there is not only one atomic limit that serves as a trivial reference state, but potentially there are many \emph{distinct} ones that depend on the symmetries considered. Each of these atomic limits induces a band structure with specific irreducible representations (irreps) of their Bloch states\cite{PhysRevLett.45.1025,Michel1999,Michel2001}. The logic proceeds then by enumerating all these atomic limits and declaring as topological any band structure with its irreps that cannot be built from such atomic limits. Thus, TQC brought a real space perspective---which was always foundational for studying SPT phases---to non-interacting topological systems.

In this work we consider the TQC standpoint that topologically trivial phases of matter can be built from atomic limits to discuss 
the concept of a
\emph{Mott SPT} phase~\cite{Kivelson10,fuji2015}, and, in general, Mott phases.
Further, inspired by TQC and symmetry-based indicators~\cite{Bradlyn_TQC,Ashwin_sym} we make use of band representations of the single-particle Green's function to investigate and detect certain interacting topological phases.
For this purpose, we invoke the concept of a topological
Hamiltonian\cite{gurarieG,wang12,Wang_2013} defined via the single-particle Green's
function, which here, we calculate within cluster perturbation
theory (CPT)\cite{CPT-ValGros,gros1994,PhysRevLett.84.522,CPT-Sene,Manghi_2013}.
We demonstrate these ideas on the example of
a Hubbard model
of spinful fermions on a diamond chain for which
we determine the phase diagram 
via infinite density matrix renormalization group (iDMRG) 
calculations~\cite{mcculloch2008,zaletel2013}
complemented by variational Monte Carlo (VMC) 
simulations.~\cite{becca2017,capello2005,capello2006,tocchio2014}
Calculations of many-body wave function topological invariants
 based on partial reflection operations\cite{Pollmann_2012} 
 serve as a reference
 for testing the Green's function--based topological classification of phases.

The paper is organized as follows:
In Section \ref{sec:HDC}, we introduce the Hubbard diamond chain Hamiltonian that we use as a testbed model to study the applicability of TQC to interacting systems. In the example of the non-interacting case, we review in Section \ref{sec:HDC_nonint} the notion of topological classification through elementary band representations (EBR) rooted in TQC.
 
In Section \ref{sec:HDC_int}, we determine the  phase diagram for the Hubbard diamond chain at half-filling via iDMRG and VMC for various values of the onsite Hubbard interaction and explore the appearance of a Mott SPT phase. We conclude this section with the computation of many-body topological invariants for the groundstate wave-functions that identify the topological nature of the interacting phases. Section \ref{sec:CPT and TH} presents the classification of topological phases through an EBR analysis of the single-particle Green's functions of the interacting system. Here, we make use of the concept of topological Hamiltonian that we combine with CPT to obtain our results. Finally, in Section \ref{sec:con} we present our conclusions and outlook.

\section{Hubbard diamond chain (HDC)}
\label{sec:HDC}
\begin{figure}
	\centering
	\includegraphics[width=\linewidth]{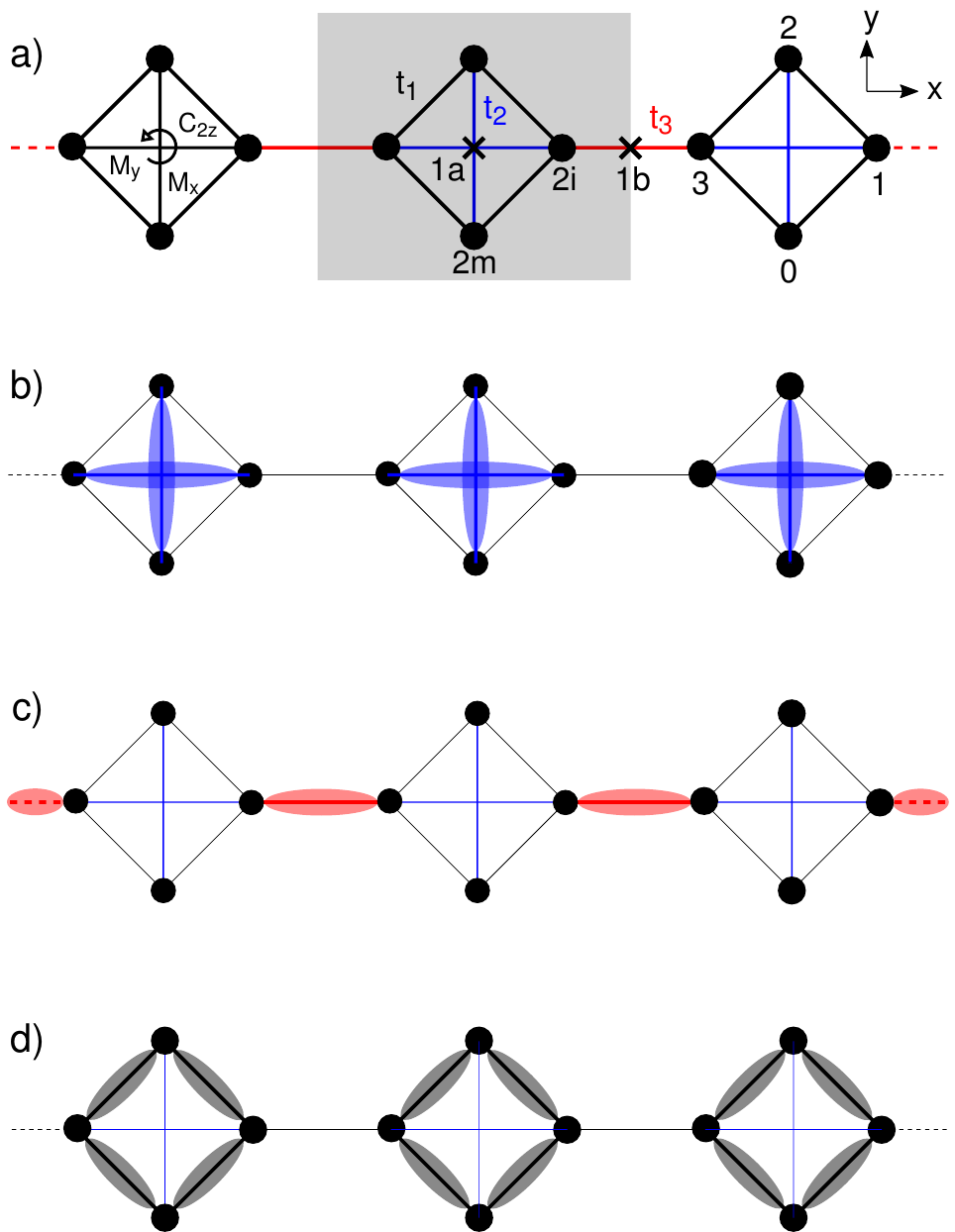}
	\caption{a) Atomic configuration of the diamond chain. The unit cell is marked with the grey background, WPs 1$a$ and 1$b$ are denoted with crosses and atomic sites at WPs 2$i$ and 2$m$ with solid black circles. The enumeration of orbitals adopted to write the Hamiltonian is also shown. Intracell hoppings $t_{1}$ and $t_{2}$ are indicated in black and blue lines,	respectively, while intercell hopping $t_{3}$ is indicated in red. Black lines and circular arrow in the left diamond denote reflection planes and 2-fold rotation with respect to the z-axis, respectively.
	b)-d) Dominant coupling parameters for three different limiting cases: 
(b) $t_2 \gg t_1, t_3$, (c) $t_3 \gg t_1,t_2$, (d) $t_{1} \gg t_{2}, t_{3}$.}
	\label{fig:model_structure}
\end{figure}

As illustrated in Fig.~\ref{fig:model_structure}a, the HDC consists of 
a one-dimensional periodic arrangement (along $x$) of diamonds with symmetry
described by the
space group $Pmmm$
(No. 47),  where the lattice sites are at  2$i$ and 2$m$ Wyckoff
positions (WPs).
We consider $s$ orbitals that induce eight bands in
reciprocal space, which have pairwise Kramers' degeneracies (spinful fermions).   

The model has three different hopping parameters: (i) an intracell nearest-neighbor hopping $t_{1}$, (ii) an intracell next-nearest-neighbor coupling $t_{2}$, and (iii) an intercell coupling $t_{3}$. Onsite electron-electron correlations are
included through a Hubbard term whose strength is controlled by the Hubbard parameter $U$. The full Hamiltonian in the absence of spin-orbit coupling is given by
\begin{align} 
\label{eq:Ham_HN}
\mathcal H=&\;U\sum_{\alpha,j}n_{\alpha,j,\uparrow}n_{\alpha,j,\downarrow} +\sum_{j,\sigma}\sum_{\alpha, \beta}c^{\dagger}_{\alpha,j,\sigma} \ \mathbb T_{\alpha\beta} \ c^{\phantom{\dagger}}_{\beta,j,\sigma}\nonumber \\
&\;-\sum_{\sigma, j}\left( t_{3} \ c^{\dagger}_{1,j,\sigma}c^{\phantom{\dagger}}_{3,j+1,\sigma} + \mathrm{h.c.}\right)
+ \mu \sum_{\alpha, j, \sigma} c_{\alpha, j, \sigma}^{\dagger} c_{\alpha, j, \sigma},
\end{align}
where $c^{\dagger}_{\alpha,j,\sigma}$ ($c_{\alpha,j,\sigma}$)
creates (annihilates) an electron of spin $\sigma$ at site $\alpha\in \{0,1,2,3\}$ of the cell labeled by $j=1,\ldots, N$ with $N$ the number of unit cells and $\mu$ is the chemical potential, which at $T=0$ matches the Fermi-energy and is chosen such that the system is at half-filling. The matrix $\mathbb{T}_{\alpha\beta}$ that contains the intracell couplings $t_{1}$ and $t_{2}$  has the form
\begin{equation}\label{eq:Hop_diamond}
\mathbb T_{\alpha \beta}=-\begin{bmatrix}
0 & t_1 & t_2 & t_1 \\
t_1 & 0 & t_1 & t_2 \\
t_2 & t_1 & 0 & t_1 \\
t_1 & t_2 & t_1 & 0  
\end{bmatrix}.
\end{equation}
Actually, the HDC model can be understood as a one-dimensional
version of the two-dimensional square lattice considered by Yao and Kivelson~\cite{Kivelson10},
where the atomic units are Hubbard diamonds.

\section{Non-interacting HDC}
\label{sec:HDC_nonint}

In this section, we study the topological nature of the non-interacting HDC within the framework of TQC. For that, we use elementary band representations of the single-valued group $Pmmm$ (the use of the single-valued group is justified by the absence of spin-orbit coupling in the Hamiltonian) to analyze the symmetry representation of bands in each phase. We will follow the notation of the {\it Bilbao Crystallographic Server}\cite{BCS1,BCS2}.

Our model is induced from  orbitals  transforming  under  the $A_{1}$ representation of the point group~\cite{C2v_vs_D2h} $C_{2v}$  (see Fig.~\ref{fig:model_structure}a) on  the  $2m$ site, and a second set of orbitals transforming under the same representation  on  the  $2i$ site.  The  8  bands  in  our  model thus  transform  under  the  composite  band  representation $(A_{1}\uparrow G)_{2i} \oplus (A_{1}\uparrow G)_{2m}$. The representations of little groups $G_{k}$ at high symmetry points $\Gamma$ and X of the BZ subduced by this representation can be decomposed as $2\Gamma_{1}^{+} \oplus \Gamma_{3}^{-} \oplus \Gamma_{4}^{-}$ and $2X_{1}^{+} \oplus X_{3}^{-} \oplus X_{4}^{-}$ in terms of irreps.

\begin{figure}
    \centering
    \includegraphics[width=0.7\linewidth]{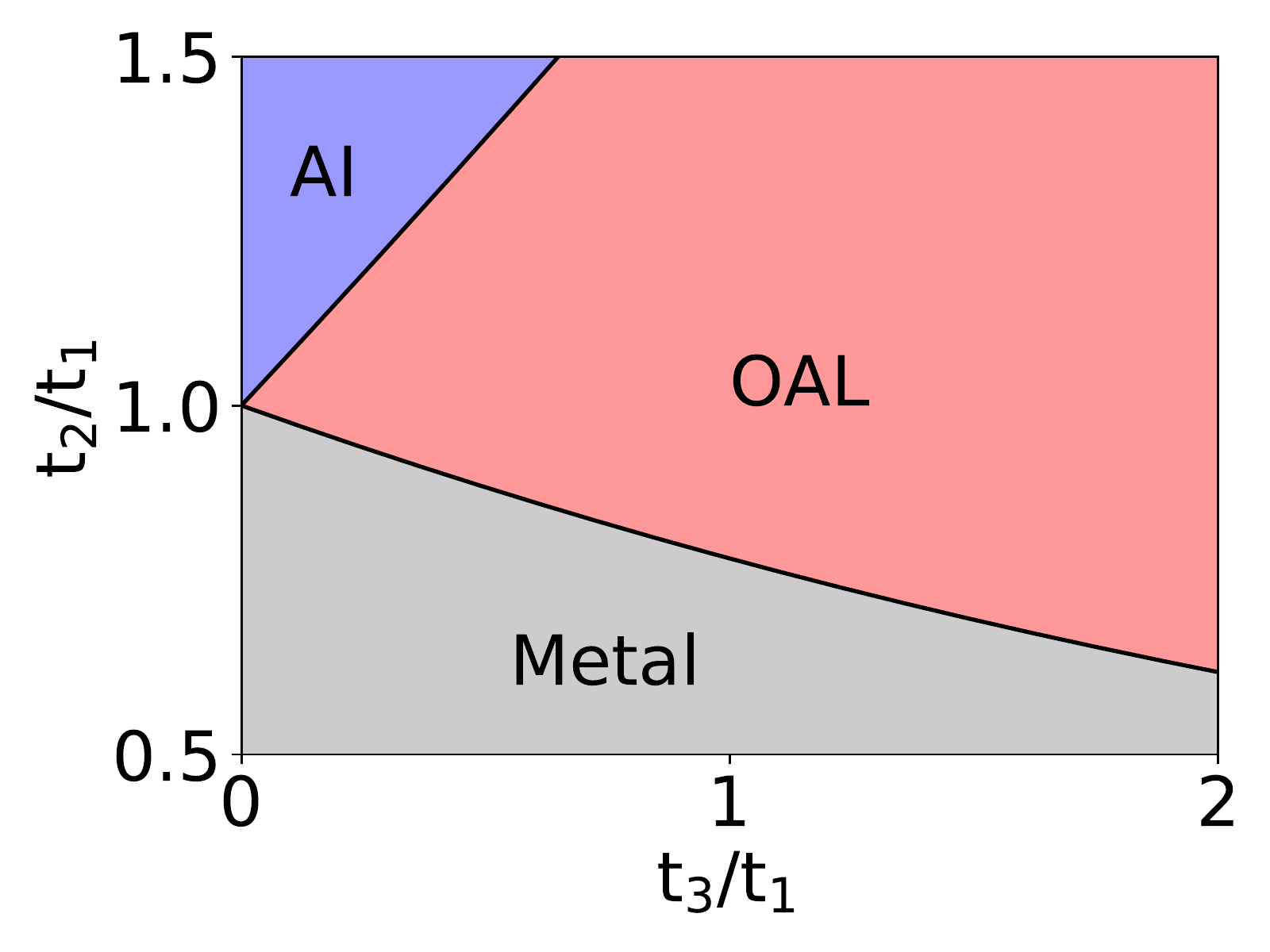}
    \caption{Phase diagram of the non-interacting diamond chain model at half-filling, the blue color represents the AI phase, red the OAL and grey the metallic phase.
    }
    \label{fig:phasediag_U=0}
\end{figure}

The analytical phase diagram of the non-interacting ($U=0$) HDC Hamiltonian~\eqref{eq:Ham_HN}  for positive $t_2/t_1$, $t_3/t_1$~\cite{pos_hop} at half-filling is shown in Fig.~\ref{fig:phasediag_U=0}. The system has a metallic and two insulating phases (labelled Metal, AI, and OAL, respectively). 

\begin{figure*}
    \centering
    \includegraphics[width=1.0\linewidth]{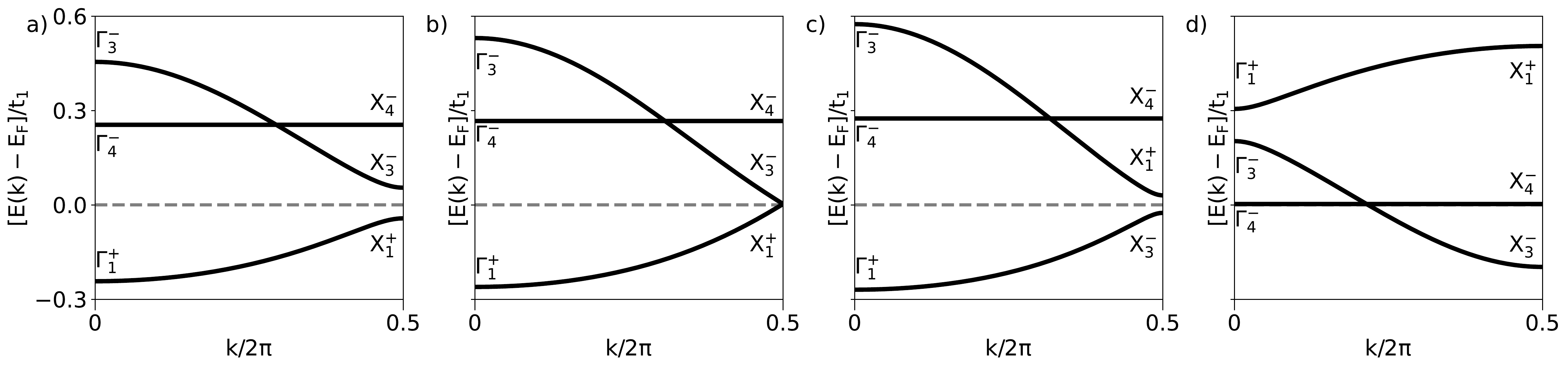}
    \caption{Band structure of the HDC for $U=0$, where the lowest occupied
	band has been omitted as it is disconnected from the rest and it
	contains the irreps $\{ \Gamma_{1}^{+}, X_{1}^{+} \}$ in the studied
	range of parameters\cite{lowestband}. (a) AI phase ($t_{2}/t_{1}=1.2$,
	$t_{3}/t_{1}=0.2$). (b) Transition point between the AI and OAL phases
	($t_{2}/t_{1}=1.2$, $t_{3}/t_{1}=0.264$). (c) OAL phase
	($t_{2}/t_{1}=1.2$, $t_{3}/t_{1}=0.3$). (d) Metallic phase
	($t_{2}/t_{1}=0.8$, $t_{3}/t_{1}=0.2$). } \label{fig:tb_bands}
\end{figure*}

In the limit $t_1 \to 0$, sites at WP $2i$ and WP $2m$ are decoupled. The sites at WP $2m$ form local dimers, while sites at WP $2i$ are connected along the periodic $x$ direction and form a one-dimensional chain that can be adiabatically connected to the Su-Schrieffer-Heeger (SSH) chain. Particularly, when $ t_3 \ll t_2 $, sites at WP $2i$ form a chain that can be connected to the trivial SSH chain. This mapping is corroborated by the TQC based analysis of the band structure: The valence bands transform in the composite band representation $2(A_{1}\uparrow G)_{1a}$, with occupied little group representations $2\Gamma_{1}^{+}$ and $2X_{1}^{+}$ (Fig.~\ref{fig:tb_bands}a). This representation can be induced from two Wannier  functions whose charge centers are at the WP $1a$ and transform like s-orbitals under the action of the site-symmetry group of this site. Since the band representation of the occupied band of the trivial SSH-chain can also be induced by identical Wannier functions, we conclude that the occupied subspace of the diamond chain's spectrum can be adiabatically connected to two copies of the trivial SSH-chain and we classify this phase as an \textit{atomic insulator} (AI).

When $t_{3}$ is the dominant hopping term, i.e. $t_{3} \gg t_1, t_2$ (see Figs.~\ref{fig:model_structure}c and \ref{fig:phasediag_U=0}), the chain formed by sites at WP $2i$ can be mapped to the SSH chain in the topological phase. Thus we expect the sites at WP 2i to contribute to the occupied subspace with a band of the same nature. Again, this connection is confirmed from the viewpoint of TQC framework: the valence bands transform in the ${(A_{1}\uparrow G)_{1a} \oplus (A_{1}\uparrow G)_{1b}}$ band representation with little group representations $2\Gamma_{1}^{+}$ and  $X_{1}^{+} \oplus X_{3}^{-}$. Particularly, the band with little group irreps $\Gamma_{1}^{+}$ and $X_{3}^{-}$ can
be induced from Wannier functions whose charge-center is in the WP $1b$ and that
transform like s-orbitals under the operations of the site-symmetry group. 
The valence band of the topological SSH-chain is obtained
from the subduction of $(A_{1}\uparrow G)_{1b}$ to the space group
$P\bar{1}$ of the SSH-chain, thus the band with little group irreps
$\{\Gamma_{1}^{+}, X_{3}^{-}\}$ in the HDC model can be mapped to the valence band of the
topological SSH-chain. On the basis of this mapping and the fact that the band with irreps $\Gamma_{1}^{+}$ and $X_{3}^{-}$ is induced from an empty WP, we identify the phase at $ t_3 \gg t_2, t_1 $ as an \textit{obstructed atomic limit} (OAL).

As can be seen in Fig.~\ref{fig:tb_bands}a-c,
 while at $\Gamma$ the irreps are the same for both phases,  at $X$ the wave
function in the valence band transforms under  $X_{1}^{+}$ for the AI phase and under $X_{3}^{-}$ for the OAL. Since these irreps have different two-fold rotation $\hat{C}_{2z}$ and
reflection $\hat{M}_{x}$ [which maps a point $(x,y,z)$ to $(-x,y,z)$] symmetry eigenvalues, it is not possible to connect these phases by a path in which the gap between valence and conduction bands does not close without breaking these symmetries. 

For the last limiting case, where $t_{1}$ is the dominant hopping and $t_1 \gg t_2, t_3$ as it is shown in Fig.\ref{fig:model_structure}d, the ground-state of the single-diamond is degenerate. Since $t_{3}$ is negligible, it follows from this degeneracy that the many-body ground-state of the HDC chain is also degenerate and therefore metallic, as it is confirmed by the presence of four partially filled bands in the band structure of Fig.~\ref{fig:tb_bands}d, where the Fermi energy is pinned at the flat band (with little group representations $\Gamma_4^-$ and $X_4^-$) associated entirely with the $2m$ sites.

\section{Topology of the interacting HDC}
\label{sec:HDC_int}
\subsection{Phase diagram of HDC for finite U } 
\label{sec:DMRG}

 We determine the phase diagram of HDC for finite $U$ values via
 infinite DMRG calculations (iDMRG) as well as variational Monte Carlo (VMC) simulations. The details of the calculations are given in
 Appendices \ref{App:DMRG} and \ref{App:VMC}.
 
 \begin{figure}
    \centering
    \includegraphics[width=0.8\linewidth]{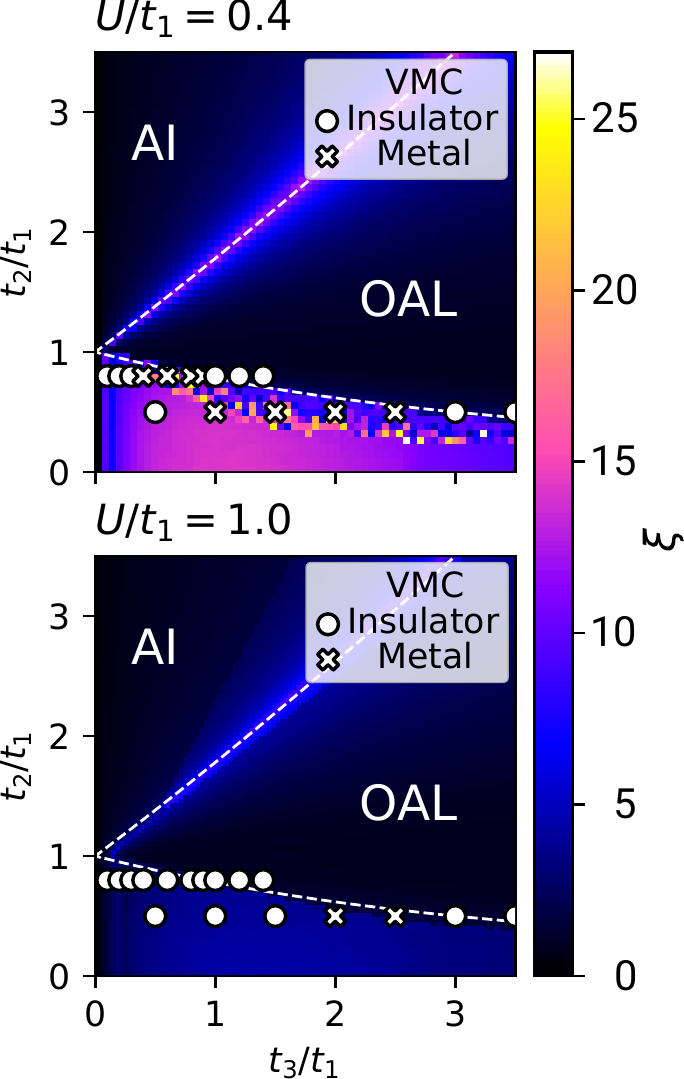}
    \caption{Phase diagram of the interacting diamond chain for $U/t_1$ = 0.4 and
    $U/t_1$ = 1 at half-filling determined from the calculation of the correlation length $\xi$
    in iDMRG (color map), as well as data points obtained by VMC at $t_2/t_1 =0.8$ and $t_2/t_1=0.5$ for various $t_3/t_1$ indicating whether the system is in a metallic (crosses) or an insulating (circles) phase. The phase boundaries of the non-interacting phase diagram are given by the white, dashed lines.}
    \label{fig:phasediag_U}
\end{figure}

  In Fig.~\ref{fig:phasediag_U}, we present the phase diagram extracted from 
 iDMRG simulations for $U/t_1$ = 0.4 and
    $U/t_1$ = 1 at half-filling by calculating the correlation length $\xi$ as defined in Eq.~\eqref{DMRG_correlation} of Appendix \ref{App:DMRG}. Phase diagrams for $U/t_1$ = 2 and
    $U/t_1$ = 4 are also shown in Appendix \ref{App:DMRG}. While gapped phases are characterized by a finite correlation length, critical points as well as metallic phases have a diverging $\xi$ \cite{hastings2006spectral}. 
 Although formally the correlation length $\xi$ diverges at the phase boundaries, it only assumes a large finite value in our data since it is bounded by the maximal bond dimension, which is set to $\chi = 128$. For metallic, or close-to-metallic systems, iDMRG performs generally poorly due to the large entanglement, resulting in points that are not fully converged close to the lower phase boundary for $U/t_1=0.4$ in Fig.~\ref{fig:phasediag_U} upper panel. 
 We have therefore performed VMC simulations as well in order to
 corroborate the iDMRG results in the region where the metallic phase is observed ($t_2/t_1 < 1$). 
 Our variational approach is based on Jastrow-Slater wave functions as described in Appendix  \ref{App:VMC}. The regions of gapped (insulator) and gapless (metallic) phases as determined by VMC are shown in Fig.~\ref{fig:phasediag_U} as circles and crosses, respectively.
 
For $t_2/t_1<1$, $t_3/t_1\ll 1 $ and any finite $U$, we find a gapped symmetry-preserving Mott-insulating (MI) phase. Increasing $t_3/t_1$, the system either undergoes a transition into an intermediate metallic phase or, for sufficiently large values of $U$, it enters the OAL phase directly from the MI phase (compare the results for $U/t_1$ = 0.4 and $U/t_1$ = 1 in the region  $t_2/t_1 = 0.8$ and $0 < t_3/t_1 < 1$ in Fig.~\ref{fig:phasediag_U}). As we elaborate in section \ref{sec:MBI}, the MI is a SPT phase that cannot be adiabatically connected to any noninteracting atomic limit provided mirror and rotation symmetries remain unbroken.  With increasing $U$, the MI replaces an increasing proportion of the metallic region of the noninteracting model, while the extent of the bordering OAL phase remains largely unchanged. 
 
For $t_2/t_1>1$ and finite interaction $U$, the transition from the AI to OAL remains but slightly shifts to larger values of $t_3/t_1$ when increasing $U/t_1$ (see also Fig.~\ref{fig:DMRG_diagrams} of Appendix~\ref{App:DMRG}). Since all single-particle bands are either completely filled or empty in the gapped AI and OAL phases at $U=0$, a small finite $U$ only induces a renormalization of the electron bands without any drastic change. In analogy with the spinful SSH chain~\cite{pollmann2012det}, both the AI and OAL phases are also smoothly connected to gapped valence bond analogues appearing at large $U$, without change of ground-state symmetry. As a result, the interacting analogues of the AI and OAL phases are each smoothly connected to a noninteracting atomic limit.

\subsection{Mott SPT phase and Many-Body Invariants}
\label{sec:MBI}

In this section, we establish the MI phase as an SPT phase and analyze the topology of the interacting phases in the HDC through many-body invariants.

The MI phase may be distinguished from the rest by the properties of the many-body ground-state wavefunction $\ket{\Psi_0}$ with respect to the crystalline symmetries. $\ket{\Psi_0}$ must transform as an irrep of the space group, which cannot change without gap closure. It is therefore sufficient to consider specific points in the phase diagram to elucidate the ground-state symmetry of each phase.

Let us focus first on the AI and OAL phases. Consider an operator $c_{i}^{\dagger}$ that creates a particle in a state that belongs to the spectra of the single-particle Hamiltonian and $h \in Pmmm$, with eigenvalue $\lambda_{h,i}$. Due to time reversal symmetry (TRS) $\mathcal{T}$, the operator $\mathcal{T} c_{i}^{\dagger} \mathcal{T}^{-1}$ corresponds to an energetically degenerate eigenstate of $h$ with eigenvalue $\lambda_{h,i}^*$. In order to obtain a many-body gapped state, these two levels must be either both unoccupied or both occupied. However, the product of operators always transforms trivially: 
\begin{align}
U_{h} c_i^\dagger( \mathcal{T} c_i^\dagger \mathcal{T}^{-1}) U_{h}^{-1} = |\lambda_{h,i}|^2 \  c_i^\dagger (\mathcal{T} c_i^\dagger \mathcal{T}^{-1}),
\end{align}
since $|\lambda_{h,i}|^2 = 1$, where $U_{h}$ is the representation of $h$. In other words, any state described by a single Slater determinant transforms trivially provided all single-particle levels are either empty or fully occupied with both spin up and spin down. This condition is satisfied by all states that can be adiabatically connected to a noninteracting gapped state of spinful particles with time reversal symmetry, in particular the AI and OAL phases.

An alternative way to show that the ground-state of the AI phase transforms as the trivial representation of $Pmmm$ follows by considering the limit $t_{3} \rightarrow 0$. In this case, $\ket{\Psi_0}$ is a product state of the local ground-states of each diamond. Let us denote with $\mathcal{O}_{j}^{\dagger}$ the operator that creates the ground-state of the $j^{\mathrm{th}}$ diamond. Then:
\begin{equation} \label{eq:productstate}
\ket{\Psi_{0}} = \mathcal{O}_{1}^{\dagger} \otimes \mathcal{O}_{2}^{\dagger} \otimes ... \otimes \mathcal{O}_{N}^{\dagger} \ket{0} \equiv \bigotimes_j \mathcal{O}_j^\dagger \ket{0},
\end{equation}
where $j$ runs over all diamonds (unit cells) and $\ket{0}$ denotes the vacuum state. Since $\mathcal{O}_j^\dagger$ contain four fermionic operators, it follows that they commute on different diamonds: $[\mathcal{O}^\dagger_i, \mathcal{O}^\dagger_j] = 0$. This provides that $\ket{\Psi_0}$ transforms trivially under discrete lattice translation:
\begin{align}
    T_x \ket{\Psi_0} = \mathcal{O}_2^\dagger \otimes ... \otimes \mathcal{O}_N^\dagger \otimes \mathcal{O}_1^\dagger \ket{0} = \ket{\Psi_0},
\end{align}
where periodic boundary conditions have been assumed. 
At $ t_2/t_1 \gg 1$ and $t_3 = U= 0$, the operator $\mathcal{O}_j^\dagger$ is given by:
\begin{align}\label{eq:OAI}
\mathcal O^{\dg}_{\text{AI},j}=&\; 
\tilde{c}_{\pi,j,\ua}^{\dagger}
\tilde{c}_{\pi,j,\da}^{\dagger}
\tilde{c}_{0,j,\ua}^{\dagger}
\tilde{c}_{0,j,\da}^{\dagger}, 
\end{align}
where $\tilde{c}^{\dg}_{\ms{k},j,\sigma}=\frac{1}{2}\sum_{\alpha}e^{i\ms{k}\alpha}c^{\dg}_{\alpha,j,\sigma}$ are Fourier transformed fermionic creation operators for a single diamond in the unit cell $j$, and $\alpha$ refers to the site labels in Fig.~\ref{fig:model_structure}a. The operators $\tilde{c}_{\pi,j,\sigma}^\dagger$ and $\tilde{c}_{0,j,\sigma}^\dagger$ both independently transform as the totally symmetric representation $A_g$ of the  point group. That is to say, they commute with all the (representation) matrices $U_{g}$ ($g \in D_{2h}$) for an individual diamond.
As a result, $ \mathcal{O}_{\text{AI},j}^{\dagger}$ transforms according to the direct product $A_g \otimes A_g \otimes A_g \otimes A_g = A_g$. 

Similarly, in the OAL phase at the limit $t_1 = U = 0$ and $t_3 \gg t_2$, the ground-state is defined by:
\begin{align}\label{eq:OOAL}
    \mathcal{O}_{\text{OAL},j}^\dagger =\prod_\sigma \frac{1}{2}(c_{0,j,\sigma}^\dagger + c_{2,j,\sigma}^\dagger)(c_{3,j,\sigma}^\dagger + c_{1,j+1,\sigma}^\dagger).
\end{align}
It follows that the corresponding ground-state $\ket{\Psi_0}$ transforms trivially under all symmetries in $Pmmm$.

For the MI phase, considering $t_3 = 0, t_2/t_1 \ll 1$ and small finite $U>0$, the ground-state of the chain is described by Eq.~\eqref{eq:productstate} and the ground-state of the diamond is given by:
\begin{align}\label{eq:OMI}
    \mathcal O^{\dg}_{\text{MI},j}=&\; \frac{1}{\sqrt{2}}\left(
\tilde{c}_{\frac{\pi}{2},j,\ua}^{\dagger}
\tilde{c}_{\frac{\pi}{2},j,\da}^{\dagger} 
-\tilde{c}_{-\frac{\pi}{2},j,\ua}^{\dagger}
\tilde{c}_{-\frac{\pi}{2},j,\da}^{\dagger}\right)
\tilde{c}_{0,j,\ua}^{\dagger}
\tilde{c}_{0,j,\da}^{\dagger},
\end{align}
which is not a single Slater determinant, and transforms instead as $B_{1g}$. It is odd with respect to $180^\circ$ rotation about the x- and y-axis (denoted $\hat{C}_{2x}$ and $\hat{C}_{2y}$), as well as mirroring in the yz- and xz-planes (denoted $\hat{M}_x$ and $\hat{M}_{y}$). These symmetries act on the operators of the chain as
\begin{align}
    \hat{C}_{2x} : c_{\alpha,j,\sigma}^\dagger \longmapsto&\; i \  c_{\beta,j,-\sigma}^\dagger [A]_{\beta\alpha},
\\
\hat{C}_{2y} : c_{\alpha,j,\sigma}^\dagger \longmapsto&\; -\sigma \  c_{\beta,N-j+1,-\sigma}^\dagger [B]_{\beta\alpha},
\\
\hat{M}_x : c_{\alpha,j,\sigma}^\dagger \longmapsto&\; i \  c_{\beta,N-j+1,-\sigma}^\dagger [B]_{\beta\alpha},
\\
\hat{M}_{y} : c_{\alpha,j,\sigma}^\dagger \longmapsto&\; -\sigma \  c_{\beta,j,-\sigma}^\dagger [A]_{\beta\alpha},
\end{align}
where $\alpha,\beta$ label sites within each diamond according to Fig.~\ref{fig:model_structure}, and $N$ is the number of diamonds in the chain. The matrices $A,B$ are given by:
\begin{align}
    A = \left(\begin{array}{cccc} 
    0 & 0 & 1 & 0 \\
    0 & 1 & 0 & 0 \\
    1 & 0 & 0 & 0 \\
    0 & 0 & 0 & 1
    \end{array} \right),\\
    B = \left(\begin{array}{cccc} 
    1 & 0 & 0 & 0 \\
    0 & 0 & 0 & 1 \\
    0 & 0 & 1 & 0 \\
    0 & 1 & 0 & 0
    \end{array} \right).
\end{align}
Since each diamond is odd with respect to these transformations, the MI ground-state is an eigenstate of each operator with eigenvalue $(-1)^{N}$. As defined, when $N$ is odd, the Mott phase is distinguished from the AI and OAL because its ground-state is odd under all four symmetries.

Therefore, when $N$ is odd, the MI phase can be distinguished from the trivial (AI) and obstructed (OAL) phases by, e.g., the mirror reflection eigenvalue of its ground-state: the observable $\langle \hat{M}_x\rangle_{\Psi_0} = \langle \Psi_0 | \hat{M}_x | \Psi_0\rangle $.  Since odd $N$ implies one diamond being a mirror center, this is simply due to the fact that the operator $\mathcal O^{\dg}_{\text{MI}}$ is odd under mirror reflection. On the other hand, when $N$ is even the ground-state reflection eigenvalue cannot detect the MI phase. A more drastic limitation of $\langle \hat{M}_x\rangle_{\Psi_0} $ is that it cannot differentiate between the AI and the OAL phase (the same is true for  $\langle \hat{M}_y\rangle_{\Psi_0} $).

In order to distinguish the AI and OAL phases, it has been proposed in Refs.~\onlinecite{Pollmann_2012, Shiozaki_2017, Shiozaki_Manybody_2017} that the ground-state eigenvalues of partial mirror reflection operations may serve as a useful diagnostic of interacting crystalline topological phases with mirror symmetry (see Appendix~\ref{ssh_invariants} for details on the case of the SSH model). Many-body topological invariants for more general point-group symmetries with a combination of internal and/or Altland Zirnbauer symmetries have also been studied within a similar framework. Here, we denote a partial reflection operation twisted by ${U}(1)$ symmetry as $\hat{M}_{x;I}(\theta)$. Such an operator has a non-trivial action on a restricted interval $I$ which contains the sites from $j=1$ to $j=L$ and average total $U(1)$ charge $Q_{I}$. It acts as
\begin{align}
\hat{M}_{x;I}(\theta): c^{\dg}_{\alpha, j,\sigma}\longmapsto&\;  i e^{-i\theta}c^{\dg}_{\beta, L-j+1,-\sigma}\left[B\right]_{\beta\alpha},
\label{eq:rt_action}
\end{align}
for $j\in I$ and trivially otherwise. We note that the partial symmetry operator $\hat{M}_{x;I}(\theta)$ generally does not commute with the Hamiltonian, so the expectation value $ \langle \hat{M}_{x;I}(\theta)\rangle_{\Psi_0}$ may evolve continuously within a given phase. Nonetheless, it is instructive to consider the limiting cases defined by Eq.~(\ref{eq:productstate}), together with Eqs.~\eqref{eq:OAI},\eqref{eq:OOAL}, and \eqref{eq:OMI}.
The results are summarized in Table~\ref{Many_body_invariants} (see App.~\ref{ssh_invariants} for details).

\begin{table}[t]
\begin{tabular}{|| p{1.2cm}|p{2.0cm}|p{4.4cm}||}
\hline
& \vspace{.5pt} \qquad $\langle \hat{M}_{x}\rangle_{\Psi_0}$ \vspace{6pt} &   \vspace{.5pt}\qquad $\langle \hat{M}_{x;I}(\theta)\rangle_{\Psi_0}$   \\
\hline \vspace{1pt}
\quad AI \vspace{5pt} & \vspace{.5pt} \hspace{.8cm} $1^{\phantom{N}}$   & \vspace{.5pt} $ \quad \ \exp\left\{-iQ_{I}(\theta-\frac{\pi}{2})\right\}$  \\
\hline \vspace{1pt}
 \hspace{2pt} OAL \vspace{5pt}& \vspace{.5pt} \hspace{.8cm} $1^{\phantom{N}}$   &  \vspace{.5pt} $ \quad \ \frac{1}{4}\exp\left\{-iQ_{I}(\theta-\frac{\pi}{2})\right\}\cos^{2}\theta$  \\
 \hline \vspace{1pt}
 \hspace{2pt}  \ MI\vspace{5pt} & \vspace{.5pt} $\ \ \   \quad (-1)^{{N}}$  & \vspace{.5pt} $ \quad \ \exp\left\{-iQ_{I}(\theta-\frac{\pi}{2})\right\}(-1)^{L}$  \\
\hline
\end{tabular}
\caption{Expectation values of the reflection operator and $U(1)$-twisted partial reflection operator evaluated for the limiting cases defined by Eq.~(\ref{eq:productstate}), together with Eqs.~(\ref{eq:OAI}), (\ref{eq:OMI}), (\ref{eq:OOAL}) for the three gapped phases of the Hubbard diamond chain model: the atomic insulator (AI), obstructed atomic limit (OAL) and the Mott insulator (MI). $Q_{I}$ denotes the average total charge enclosed within the interval $I$.} 
\label{Many_body_invariants}
\end{table}

As can be seen from Table~\ref{Many_body_invariants}, the partial mirror reflection operator provides a sharper diagnostic to detect and distinguish all three phases. The factor of $e^{-iQ_{I}\theta}$ is common to all the phases and simply detects the $U(1)$ charge $Q_{I} $ enclosed within the interval $I$. The three phases can in particular be distinguished by choosing $\theta=\pi/2$, and $L$ an odd integer. For this choice and small $U$, the limiting values of the topological indicator are 1,0 and -1 (since $Q_I$ is a multiple of 4) for the AI, OAL and MI phase respectively. If the same hoppings are considered with large $U$, these become 1, $\frac{1}{2}$, and $-1$, instead. Away from these ideal limits, the topological indicators are expected to remain close to the ideal values provided the correlation length remains short \cite{Pollmann_2012, Shiozaki_2017, Shiozaki_Manybody_2017}, allowing them to function as a diagnostic of the ground-state topology. 

The MI phase in the HDC chain is not adiabatically connected to {\it any} non-interacting atomic limit, yet it can be continued to a state with no entanglement between the unit cells (the limit $t_3\to 0$). A similar case was illuminated in Refs.~\onlinecite{Kivelson10,fuji2015}. The Hubbard diamond thus enriches the possible building blocks of quantum matter. This observation necessitates the expansion of possible atomic limits to include {\it Mott} or {\it interacting atomic limits}.

\section{Topology and Green's functions}
\label{sec:CPT and TH}
We explore now to which extent the topology of the interacting phases in the HDC model  can be identified by using eigenstate representations of the single-particle Green's function  inspired by TQC. For that we make use of the concept of a topological Hamiltonian and apply CPT to calculate the Green's functions for the interacting system.

\subsection{Topological Hamiltonian}
\label{sec:TH}

We first shortly review the concept of the topological Hamiltonian which allows to define topological invariants in terms of the single-particle Green's function in an interacting system.~\cite{Volovik2003,gurarieG,wang10,wang12inv,wang12,Wang_2013,mertz2019}

In Ref.~\onlinecite{wang12} it was realized that it is sufficient to focus on the Green's function at zero frequency to obtain topological invariants. Equivalently, it is possible to define an auxiliary non-interacting Hamiltonian, denoted topological Hamiltonian, from which the Green's function invariants can be calculated

\begin{equation} \label{eq:def_Htopo}
H_{\textrm{T}}(\mathbf{k}) = -G^{-1}(0, \mathbf{k}).
\end{equation}
The topological Hamiltonian $H_{\textrm{T}}$ is hermitian and it is well-defined
as long as
there is a gap in the spectral function around zero frequency and $G(0,
\mathbf{k})$ does not have a zero eigenvalue. Furthermore the
topological Hamiltonian possesses the same spatial symmetries as the
interacting many-body Hamiltonian under the assumption that the many-body
ground-state is unique.

Under these conditions it is possible to generalize the formalism of TQC and
symmetry-based indicators to study the Green's functions in terms of the topological
Hamiltonian~ \cite{Lessnich2021}. 
Symmetry representations of valence bands and symmetry indicators can directly be computed from the topological Hamiltonian and they can only change if either (i) the gap closes in
the spectral function at zero frequency, (ii) the Green's function has a zero eigenvalue at zero frequency, or (iii) the Green's function breaks a protecting symmetry.

In what follows, we apply this formalism to our testbed HDC, where the Green's functions are obtained from CPT. 
With this analysis, we can gain insights into the correspondence between the topological characterization of the ground-state for the interacting system presented in Section \ref{sec:HDC_int} and the topological characterization  performed in terms of Green's functions.

We emphasize that there is a difference between investigating the adiabatic connectedness of the ground-state of an insulator and topological invariants defined in terms of the Green's function when interactions are present (see also the discussion in Refs.~\onlinecite{gurarie1d,limitG2,unkonv_pert}).

\subsection{CPT and topological Hamiltonian for HDC}\
\label{sec:CPT-HT}

We use CPT to calculate the Green's function of the interacting HDC model.
CPT is a numerical technique for calculating the
Green's functions of strongly-correlated electrons described by Hubbard models
in periodic lattices~\cite{CPT-ValGros,gros1994,CPT-Sene}. The basic idea behind CPT is
to divide the lattice into a (super)lattice of clusters. The Hubbard model on 
each cluster is solved exactly, whereas hoppings between sites belonging to
different clusters are treated perturbatively. Our choice of cluster is the 4-site diamond (grey region in Fig.~\ref{fig:model_structure}a).
Details of the method and calculations of the single-particle Green's function $G(\omega, \bs{k})$ [and spectral function
$A(\omega, \bs{k})$] for the Hubbard diamond chain 
are given in Appendix~\ref{App:CPT}. 

\begin{figure*}
    \centering
    \includegraphics[width=1.0\linewidth]{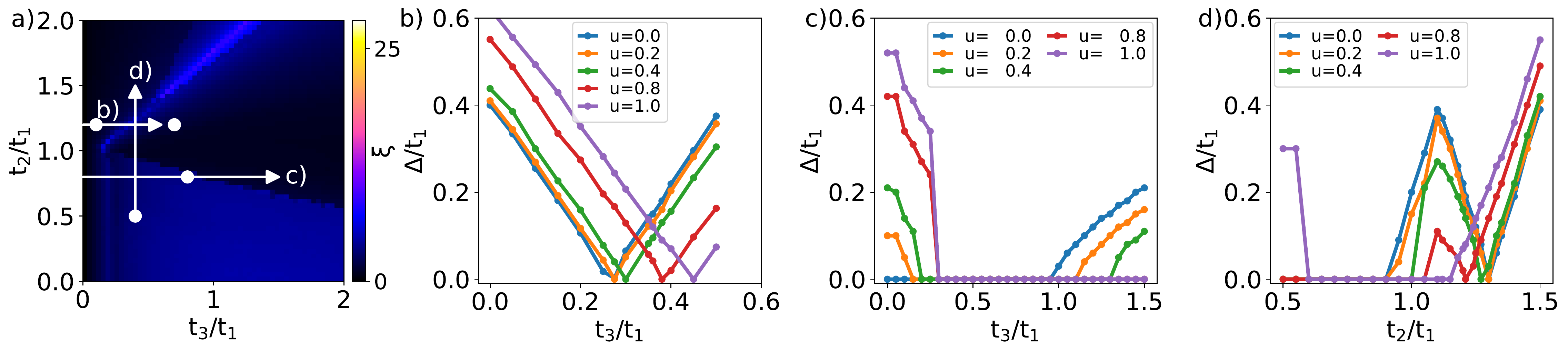}
    \caption{(a) Labeling of the paths considered in the CPT calculations over the iDMRG/VMC phase diagram obtained with $U/t_{1}=1.0$. (b), (c) and (d) show the evolution of the 
    charge gap $\Delta/t_1$ along the different paths as a function of the hopping parameter for different values of the $u=U/t_1$. A cluster containing a single unit cell has been used in CPT calculations.}
    \label{fig:CPT_gaps}
\end{figure*}

\begin{figure}
    \centering
    \includegraphics[width=1.0\linewidth]{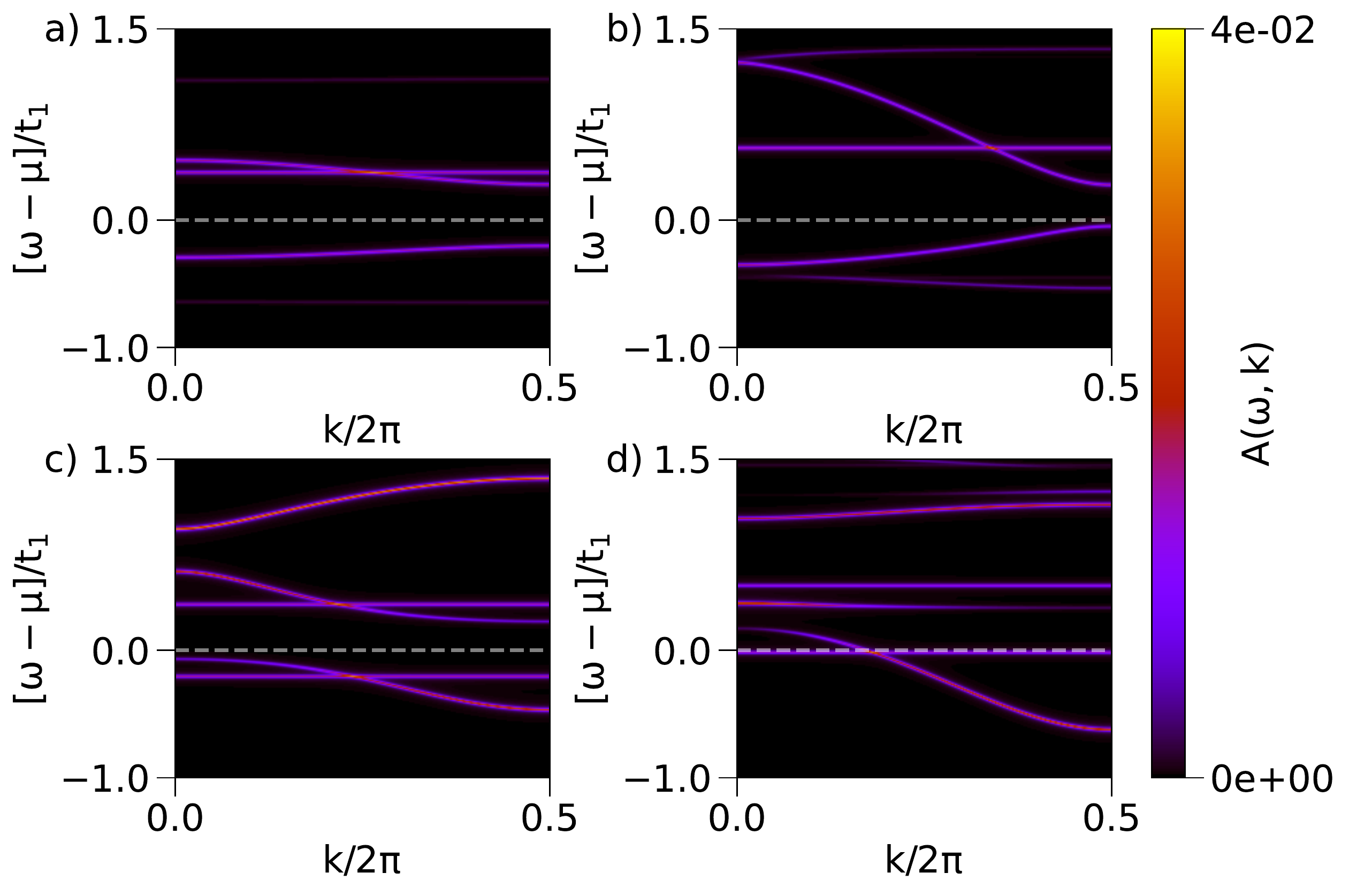}
    \caption{CPT spectral functions  $A(\omega, k)$ calculated for $U/t_1=1$ at the marked cicles in Fig.~\ref{fig:CPT_gaps}a. The Fermi-level is shown with dashed lines. Each spectral function is normalized to satisfy $\int \mathrm{d}\omega\sum_{k} A(\omega, k)=1$, where the integral runs over the whole frequency domain. (a) AI ($t_{2}/t_{1}=1.2$, $t_{3}/t_{1}=0.1$), (b) OAL ($t_{2}/t_{1}=1.2$, $t_{3}/t_{1}=0.7$) , (c) MI ($t_{2}/t_{1}=0.5$, $t_{3}/t_{1}=0.4$), (d) Metallic phase ($t_{2}/t_{1}=0.8$, $t_{3}/t_{1}=0.8$).}
    \label{fig:CPT_awk}
\end{figure}

We first check the reliability of the CPT results for the interacting HDC at half-filling.
For that we show in  Fig.~\ref{fig:CPT_gaps}b-d the calculated  charge gaps $\Delta/t_1$
[extracted from the spectral function $A(\omega, k)$] at 
various values of interaction strength $U$ and along three different hopping paths in the phase diagram as marked 
in Fig.~\ref{fig:CPT_gaps}a.
We compare the results with the phase diagram 
obtained from  iDMRG and VMC in Fig.~\ref{fig:phasediag_U}.
We identify four phases: three insulators characterized by the presence of a charge-gap in the spectral function and a metallic phase, in agreement with iDMRG and VMC. Spectral functions calculated at representative points
in each of these phases  (marked with circles in Fig.~\ref{fig:CPT_gaps}a) are shown in Fig.~\ref{fig:CPT_awk}.

The insulating phase at finite $U$, 
	$t_{2}/t_{1}<1$ and $t_{3}/t_{1} \ll 1$  corresponds to the Mott phase
	where we identify the origin of the charge gap through the formation
of upper and lower Hubbard bands (see Fig.~\ref{fig:CPT_awk}c).
The remaining two insulating phases at $t_2/t_1>1$ are reminiscent of the non-interacting AI
and OAL phases. The corresponding spectral functions suggest that they
are correlated insulators (see
Figs.~\ref{fig:CPT_awk}a-b). As $U$ increases, the phase transition
connecting both phases is shifted towards larger values of $t_{3}/t_{1}$ (Fig.\ref{fig:CPT_gaps}b), which is
in good agreement with the iDMRG calculations shown in
Fig.~\ref{fig:phasediag_U}.

At $t_{2}/t_{1} < 1$, and moderate values of $U$ a metallic phase appears  between the Mott insulator and one of the correlated insulating phases in a narrow region of intermediate $t_{3}/t_{1}$ values (see Fig.\ref{fig:CPT_gaps}c) in agreement with iDMRG and VMC.
Within the range of $U$ values considered in our calculations,
we observe that an increase of  $U$  shifts the
gap closing (opening) that indicates the onset (offset) of the metal phase to
larger values of $t_{3}/t_{1}$. Since the choice of the cluster adopted for the
CPT calculations treats the intercluster-hopping $t_{3}$ as a perturbation,  
we expect a loss in the CPT performance at large $t_{3}/t_{1}$ and therefore, in that region the results are less reliable.

\begin{figure}
    \centering
    \includegraphics[width=1.0\linewidth]{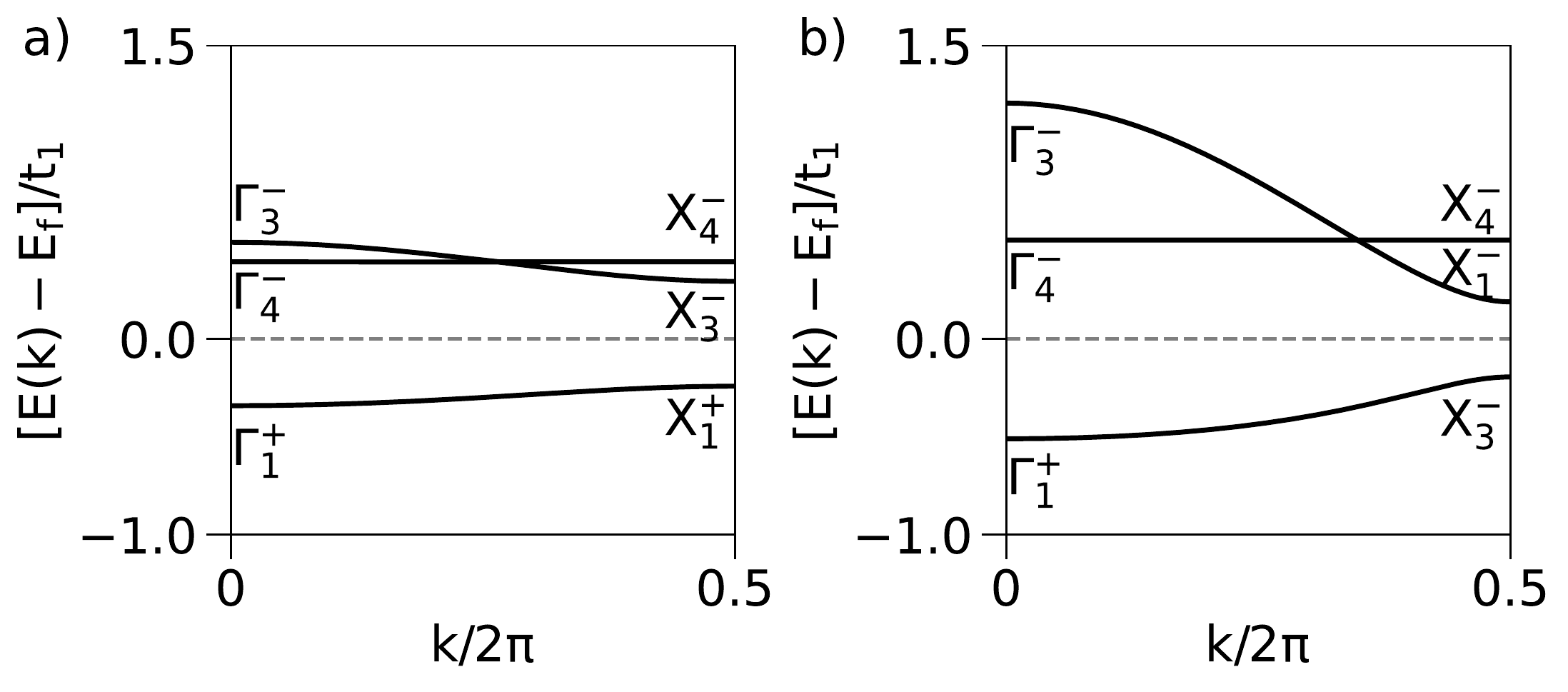}
    \caption{Spectrum and irreps of the topological Hamiltonian in the correlated insulating phases for $U/t_1=1$ in a) the AI phase ($t_{2}/t_{1}=1.2$, $t_{3}/t_{1}=0.1$) and b) the OAL phase ($t_{2}/t_{1}=1.2$, $t_{3}/t_{1}=0.7$). The lowest occupied band has been omitted\cite{lowestband}}
    \label{fig:CPT_ht_bands}
\end{figure}

With the calculated Green's functions we construct the topological Hamiltonian and proceed with the determination of the topological nature of the
insulating phases appearing  at $t_{2}/t_{1} > 1$ by 
analyzing the topological Hamiltonian's spectrum. The use of the topological Hamiltonian [Eq.~\eqref{eq:def_Htopo}] in this region is justified by
the fact that $G(0, k)$ is non-singular for these insulating phases.
In Fig.~\ref{fig:CPT_ht_bands}, we show the spectrum of the topological Hamiltonian for both 
phases at  $U/t_1$=1 together with the irreps of the little groups at $\Gamma$ and $X$ under which the topological Hamiltonian's bands transform. 
We note that the irreps of the occupied bands at small (large) values of
$t_{3}/t_{1}$ (Fig.~\ref{fig:CPT_ht_bands}a and b, respectively) coincide with those of the
non-interacting AI (OAL) phases (compare with Fig.~\ref{fig:tb_bands}a and c, respectively), which suggests that these interacting insulating phases are adiabatically connected
to the non-interacting ones. Likewise the boundary between both AI and OAL phases is through a (charge) gap closing. 

\begin{figure}
    \centering
    \includegraphics[width=0.9\linewidth]{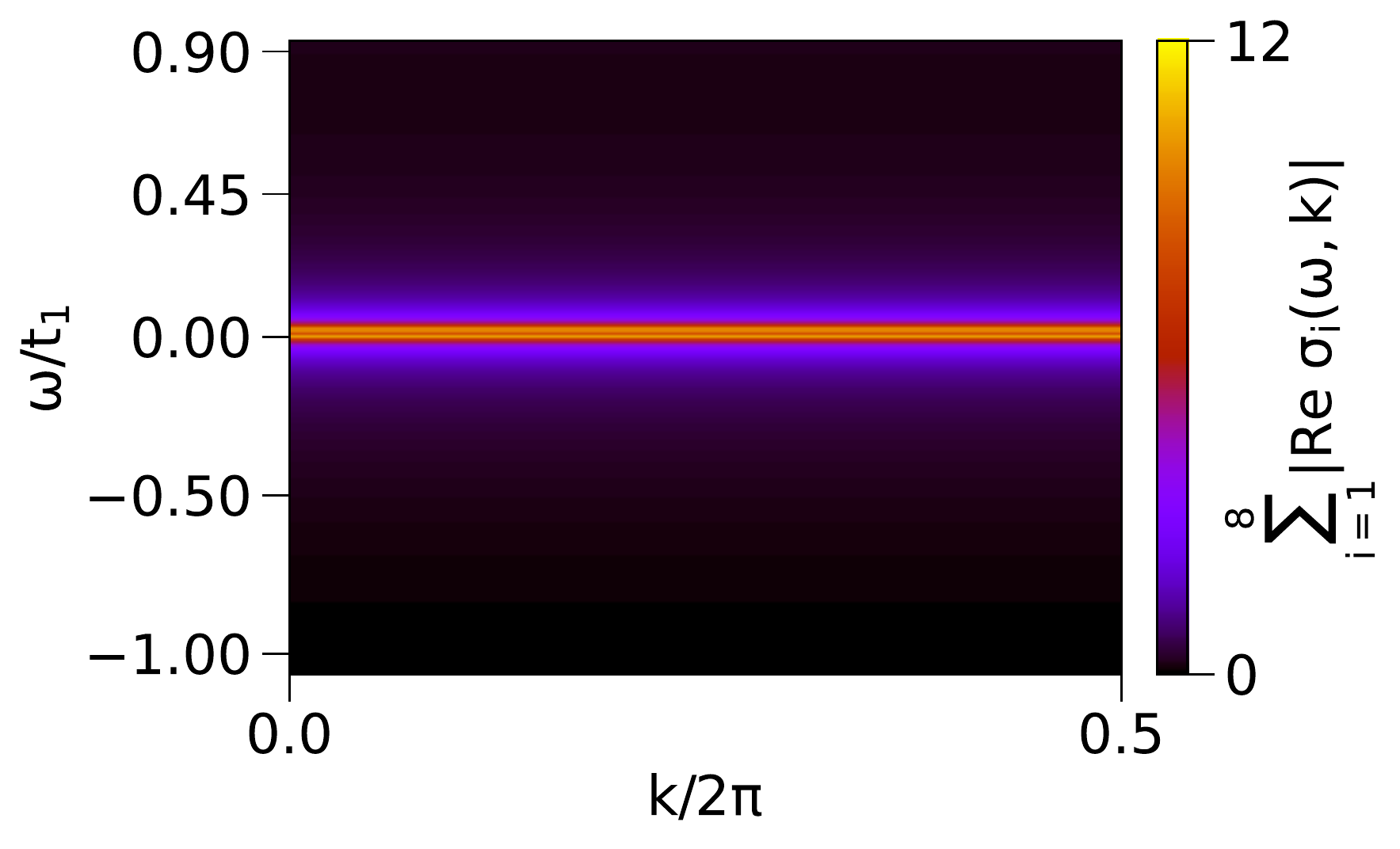}
    \caption{Sum of the absolute value of the real part of self-energy's eigenvalues $\sigma_{i}(\omega, k)$ in the Mott SPT phase. The drastic increase close to $\omega=0$ is identified as a divergence of the self-energy.}
    \label{fig:self-energy_diver}
\end{figure}

The Mott SPT phase, on the contrary, deserves special attention. In section~\ref{sec:MBI} it was shown that there is no smoothly connected non-interacting limit to it. A calculation of CPT Green's functions in this phase indicates that $G(0, k)$ is singular. This can be monitored by a diverging self-energy as displayed in Fig.~\ref{fig:self-energy_diver} for $t_2/t_1 = 0.5$, $t_3/t_1 =0.4$ 
and $U/t_1 = 1.0$. In such a situation, the topological Hamiltonian is not applicable and the EBR description cannot be pursued. 

With this analysis we conclude that the combination of topological Hamiltonian and TQC provides a possible new avenue to characterize correlated
insulating phases as far as they are smoothly connected to non-interacting limits. This excludes Mott phases which require a characterization beyond the single-particle Green's functions.

\section{Conclusions and Outlook} 
\label{sec:con}

In this work, we have explored the possibility of extending the TQC formalism
	to correlated systems by studying the specific example of a Hubbard diamond chain.	After determining the phase
	diagram of the  model using infinite density matrix renormalization group calculations and variational Monte Carlo	simulations, we investigated the topology of all phases with  many-body topological invariants. We thusly identified three insulating phases (AI, OAL and Mott SPT) and a metallic phase depending on the
	interaction strength. Specifically, we demonstrated that the Mott phase is a symmetry protected topological phase which is not adiabatically connected to any band insulator, contrary to the AI and OAL phases.
	
	Further, we investigated a generalization of the TQC formalism to Green's functions combined with the concept of topological Hamiltonians to identify the topological nature 
	of the interacting phases, using
	cluster perturbation theory to calculate the Green's functions. We illustrated that this approach provides a possible recipe to characterize the topology of interacting insulating phases, as far as they are adiabatically connected to non-interacting phases. While we focused on the example of the one-dimensional Hubbard diamond chain with space group $Pmmm$, the formalism we introduced can easily be extended to systems with other space groups and/or higher dimensions.

The topology of the Mott phase, in contrast, cannot be detected by this approach. It fundamentally requires the knowledge of $n$-particle Green's functions ($n>1$). A systematic extension of the TQC formalism to this case may be able to identify what type of Mott atomic limit a given phase corresponds to. Before such a program can be carried out, however, the representation theory of $n$-particle Green's functions needs to be developed, which we leave for future work.

\section*{ACKNOWLEDGMENTS}

We thank J.L. Ma\~{n}es, B. A. Bernevig, B. Bradlyn, J. Cano and F. Becca for fruitful discussions. M.G.V.  and  M.I.
acknowledge support from the  Spanish  Ministerio  de  Ciencia  e  Innovacion
(grants number PID2019-109905GB-C21 and PGC2018-094626-B-C21) and Basque Government (grant IT979-16).
F.F. acknowledges support from the Alexander von Humboldt Foundation through
a postdoctoral Humboldt fellowship. N.H acknowledges support from the Stiftung Polytechnische Gesellschaft (SPTG, Foundation Polytechnical Society Frankfurt, Germany) through a Master's fellowship. D.L., T.M. and R.V. acknowledge the
Deutsche Forschungsgemeinschaft (DFG, German Research
Foundation) for funding through Grant No. TRR 288 -
422213477 (project B05). A.T. acknowledges funding by the European Union’s
Horizon 2020 research and innovation program under 
the Marie Sklodowska Curie grant agreement No 701647. Part of the work of M.G.V., F.P., and R.V. was
carried out at Kavli Institute of Theoretical Physics (KITP), which is
supported by  the National Science Foundation under Grant No.NSF PHY-1748958. 
T.N. and F.P. acknowledges funding from the European Union’s Horizon 2020 research and innovation programm (ERC-StG-Neupert-757867-PARATOP and ERC-CoG-Pollmann-771537-DYNACQM).

\appendix

\section{DMRG calculations}\label{App:DMRG}
The Density Matrix Renormalization Group (DMRG) algorithm is one of the most powerful and unbiased numerical methods for one-dimensional and quasi one-dimensional systems.\cite{white1992density, white1993density} Our calculations have been performed using the infinite DMRG (iDMRG) method\cite{mcculloch2008infinite}, which is an extension of standard DMRG to infinite systems, as implemented in the TeNPy package.\cite{hauschild2018efficient}

We initialize the algorithm on a two-diamond unit cell as the half-filled product state $|\Psi\rangle_0 = |\downarrow, \downarrow, \uparrow, \uparrow, \downarrow, \downarrow, \uparrow, \uparrow\rangle$, with the sites ordered as given in \mbox{Fig.\ \ref{fig:model_structure}}. From there, we build a matrix product state (MPS) representation of the form
\begin{align}
    |\Psi\rangle = \sum_{j_1\ldots j_N}M^{\left[ 1 \right]j_{1}}M^{\left[2 \right]j_{2}}\ldots M^{\left[N \right]j_{N}}|j_1, j_2, \ldots, j_N\rangle,
\end{align}
where each $M^{\left[n \right]j_{n}}$ is a $\chi_n\times \chi_{n+1}$ matrix, and $N$ the number of sites. We employ the commonly used two-site update, which sweeps through the system and iteratively optimizes the matrices by minimizing the energy locally with respect to our Hamiltonian \eqref{eq:Ham_HN}, keeping the number of electrons fixed. The procedure is repeated until the convergence criteria are fulfilled ($\Delta E < 10^{-10}$ eV and $\Delta S < 10^{-4}$ UNIT). 

\begin{figure}
    \centering
    \includegraphics[width=\linewidth]{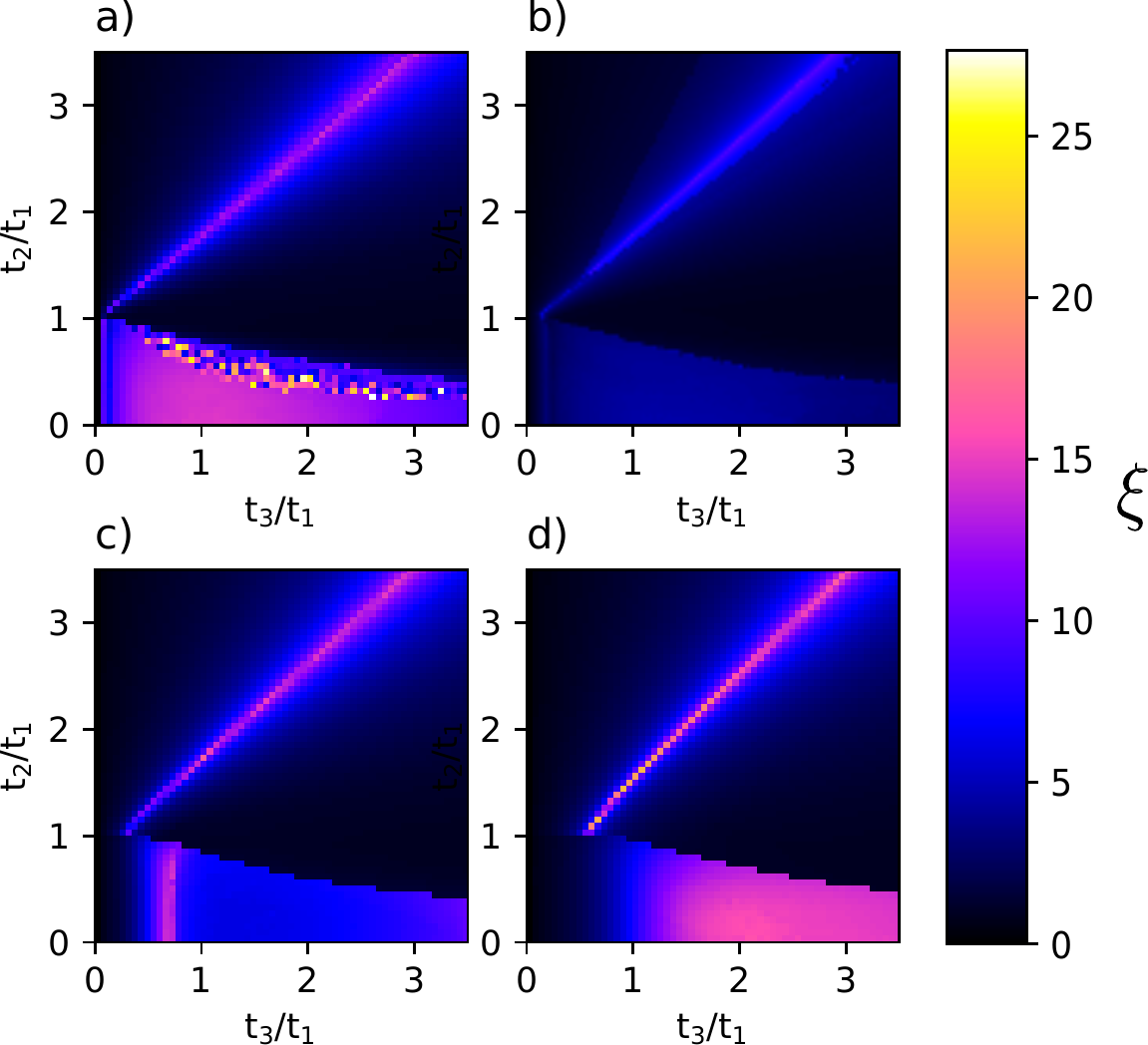}
    \caption{iDMRG calculations for the HDC model showing the correlation length $\xi$ for Hubbard interaction strengths a) $U/t_1 = 0.4$, b) $U/t_1 = 1.0$, c) $U/t_1 = 2.0$ and d) $U/t_1 = 4.0$. The maximal bond dimension is set to $\chi = 128$.}
    \label{fig:DMRG_diagrams}
\end{figure}

Having calculated the ground-state, we compute the correlation length, which, in the DMRG context, is defined as 
\begin{align}
    \xi = -\frac{N}{\log |\eta_2|},
    \label{DMRG_correlation}
\end{align}
with $\eta_2$ denoting the second largest eigenvalue of the transfer matrix $T$.\cite{hauschild2018efficient} While gapped phases are characterized by a finite correlation length, critical points, as well as metallic phases, have a diverging $\xi$.\cite{hastings2006spectral}  The resulting phase diagrams are presented in \mbox{Fig.\ \ref{fig:DMRG_diagrams}}. The non-interacting phase diagram as shown in \mbox{Fig.\ \ref{fig:phasediag_U=0}} is certainly recognizable in the DMRG results. 

Although formally, the correlation length $\xi$ diverges at the phase boundaries, it only assumes a large finite value in our data, since it is bounded by the maximal bond dimension, which is set to $\chi = 128$. For metallic, or close-to-metallic systems, DMRG performs generally poorly, resulting in points that are not fully converged close the lower phase boundary for $U=0.4$ in \mbox{Fig.\ \ref{fig:DMRG_diagrams}a)}. 

For $t_2 = 0$ and $t_3 \ll t_1$, we expect a Mott-insulating phase for any finite value of $U$. Increasing the Hubbard interaction, the short-range correlated Mott region in the lower left hand corner of the phase diagram extends further to the right. Increasing $t_3$, the system either undergoes a transition into an intermediate metallic phase or, for sufficiently large values of $U$, it enters the OAL phase directly. The exact value of $U$ at which the intermediate phase is completely suppressed is hard to pinpoint, due to the strong drift observed in the data. 

\begin{figure}
    \centering
    \includegraphics[width=\linewidth]{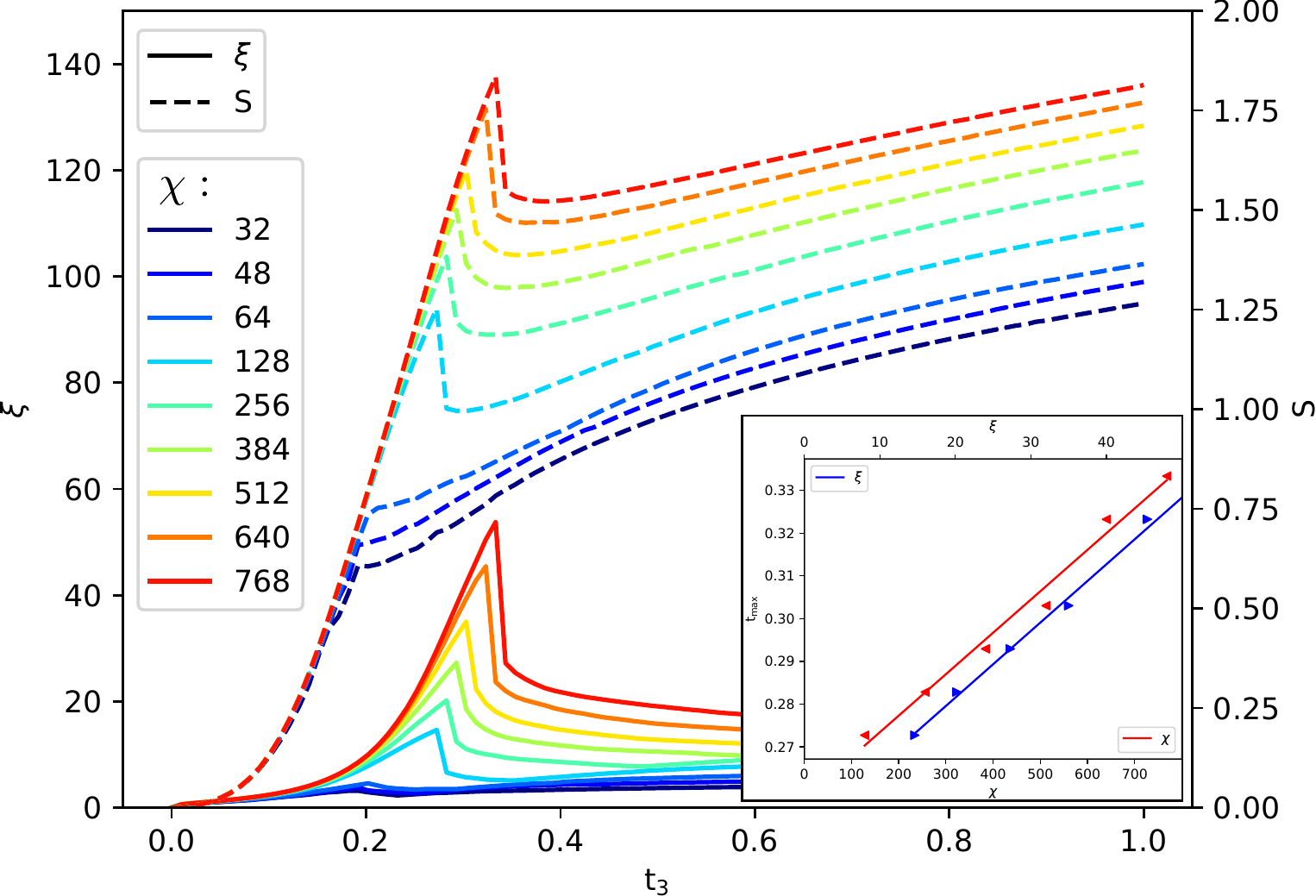}
    \caption{Correlation length $\xi$ (solid line) and entanglement entropy $S$ (dashed line) plotted against $t_3/t_1$ for $U/t_1 = 1.0$ and $t_2/t_1 = 0.3$ for different values of $\chi$. }
    \label{fig:peak_position}
\end{figure}

 Fixing $t_2/t_1 = 0.3$ and $U/t_1 = 1.0$, we plot the correlation length $\xi$ and the entanglement entropy $S$ against $t_3/t_1$ for different maximal bond dimensions $\chi$ in \mbox{Fig.\ \ref{fig:peak_position}}. With increasing $\chi$, we note that the position of the peak $t_{max}$ shifts to higher values of $t_3/t_1$ as shown in the inset of the figure. Extrapolating this behavior to infinitely large values of $\chi$ suggests that the Mott-metal transition is suppressed, and the system enters the OAL phase directly. 

Using finite DMRG, we calculate the charge gap crossing the AI-OAL phase boundary at $t_2/t_1 = 1.2$ with $U/t_1 = 1.0$ on a chain of 20 diamonds (80 sites). As in the non-interacting case, we observe a closure of the charge gap at the transition. 

\section{Variational Monte Carlo calculations}\label{App:VMC}

To strengthen our results for the phase diagram of the diamond chain, we also perform variational Monte Carlo (VMC) calculations in the region of the phase diagram in which the metallic phase is observed ($t_2/t_1<1$). Our variational approach is based on Jastrow-Slater wave functions of the form
\begin{equation}\label{eq:psivar}
|\Psi_{var}\rangle=\mathcal{J}_n \mathcal{J}_s |\Phi_0\rangle,
\end{equation}
in which long-range Jastrow correlators, $\mathcal{J}_n$ and $\mathcal{J}_s$, are applied onto an uncorrelated fermionic state, $|\Phi_0\rangle$, to introduce non-trivial electron-electron correlations. This class of variational states has been shown to accurately describe both metallic and Mott insulating phases in one dimension~\cite{capello2005}. The variational \textit{Ansatz} of Eq.~\eqref{eq:psivar} features long-range density-density and spin-spin Jastrow factors,
\begin{align}
 \mathcal{J}_n & =\exp\left(\sum_{i,j}\sum_{\alpha,\beta} 
 v_{\alpha,i;\beta,j} n_{\alpha,i} n_{\beta,j} \right), \\
 \mathcal{J}_s & =\exp\left(\sum_{i,j}\sum_{\alpha,\beta} 
 u_{\alpha,i;\beta,j} S^z_{\alpha,i} S^z_{\beta,j} \right),
\end{align}
and the non-interacting state $|\Phi_0\rangle$. Although the simplest choice for $|\Phi_0\rangle$ is the ground-state of the Hamiltonian~\eqref{eq:Ham_HN} with $U=0$, here we adopt a more general scheme, in which we consider the ground-state of an auxiliary quadratic Hamiltonian~\cite{becca2017}
\begin{align}
\mathcal{H}_{0} = & \sum_{i,j}  \sum_{\alpha,\beta} \ \biggl [ \sum_\sigma
t_{\alpha,i;\beta,j} \ c_{\alpha,i,\sigma}^\dagger c_{\beta,j,\sigma}^{\phantom{\dagger}} +H.c. \nonumber
 \\
 & + \Delta_{\alpha,i;\beta,j}  \ (c^\dagger_{\alpha,i,\uparrow} c^\dagger_{\beta,j,\downarrow} + c^\dagger_{\beta,j,\uparrow} c^\dagger_{\alpha,i,\downarrow}) + H.c. \biggr]
 \nonumber \\
 & + \Delta_{\rm AF} \sum_{j}\sum_{\alpha}
 \left [ e^{i \pi(j+\alpha)} c_{\alpha,j,\uparrow}^\dagger c_{\alpha,j,\downarrow}^{\phantom{\dagger}}
 + H.c. \right ].
\end{align}
$\mathcal{H}_0$ contains hopping terms ($t_{\alpha,i;\beta,j}$) and singlet pairing terms ($\Delta_{\alpha,i;\beta,j}$) up to fifth-neighbors, and a N\'eel magnetic field ($\Delta_{\rm AF}$). In order to minimize the variational energy of the trial state, all the parameters of $\mathcal{H}_{0}$ and the Jastrow pseudopotentials ($v_{\alpha,i;\beta,j}$, $u_{\alpha,i;\beta,j}$) are optimized by means of the stochastic reconfiguration technique~\cite{sorella2005,becca2017}.

\begin{figure}
\centering
\includegraphics[width=1.0\linewidth]{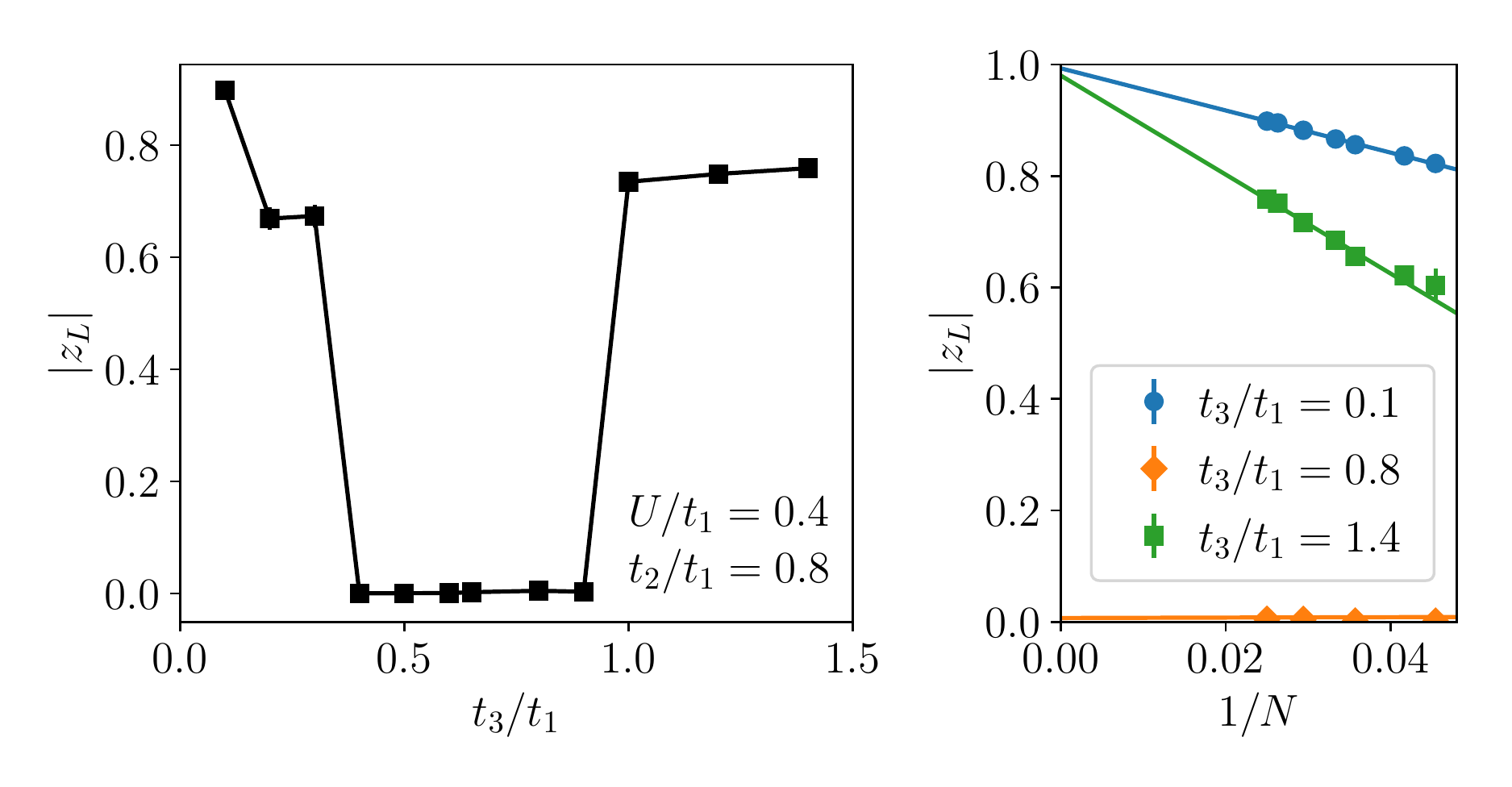}
\caption{Localization parameter $z_L$ [Eq.~\eqref{eq:zloc}] computed by variational Monte Carlo for $U/t_1=0.4$ and ${t_2/t_1=0.8}$. Left panel: $|z_L|$ for a system of $N=40$ diamonds ($160$ sites) and different values of $t_3/t_1$. We observe a metallic phase ($|z_L|\approx 0$) sandwiched between two insulating phases (finite $|z_L|$). Right panel: finite size scaling of $|z_L|$ at ${t_3/t_1=0.1} \mbox{ (insulator)},0.8\mbox{ (metal)},1.4\mbox{ (insulator)}$.}
\label{fig:zl_vmc}
\end{figure}

When scanning the phase diagram of the diamond chain, we can discriminate between metallic and insulating phases by computing two distinct observables. On the one hand, we can evaluate the density-density structure factor $N(q)=\langle n_{-q} n_q \rangle_{var}$, where ${n_q=N^{-1}\sum_{j,\alpha} n_{\alpha,j} \exp(iqj)}$ is the Fourier transform of the density operator and ${\langle \cdots \rangle_{var}}$ indicates the expectation value with respect to the variational state~\eqref{eq:psivar}. The absence (presence) of a charge gap is signalled by the linear (quadratic) behavior of $N(q)$ for $q\rightarrow 0$~\cite{feynman1954,capello2005,capello2006}. On the other hand, we can compute the expectation value of the localization parameter introduced in Ref.~\onlinecite{resta1999}, namely
\begin{equation}\label{eq:zloc}
 z_L=\left\langle\exp\left(\frac{2\pi i}{N} \sum_{j,\alpha} j n_{\alpha,j}\right)\right\rangle_{var}.
\end{equation}
In the thermodynamic limit, $|z_L|\rightarrow 0$ in a metallic phase, while $|z_L|\rightarrow 1$ in an insulating phase~\cite{capello2005} (see Fig.~\ref{fig:zl_vmc} for an example).

We performed VMC calculations for $U/t_1=0.4$ and $U/t_1=1$, $t_2/t_1=0.5$ and $t_2/t_1=0.8$, and different values of $t_3/t_1$. The results are reported in Fig.~\ref{fig:phasediag_U}, on top of the DMRG phase diagram.

\section{Benchmarking the topological invariants with the Su-Schrieffer-Heeger model}
\label{ssh_invariants}
Let us consider the fixed point Su-Schrieffer-Heeger (SSH) model described by the Hamiltonian
\begin{align}
H(\alpha)=\sum_{j=1}^{N-1}b_{j}^{\dagger}a_{j+1}+e^{-i\alpha}b_{N}^{\dg}a_{1}+\text{h.c}
\end{align}
where we have inserted a ${U}(1)$ flux by twisting the boundary conditions by $e^{-i\alpha}$. Let us define basis transformed fermions as
\begin{align}
f^{\dg}_{j+\frac{1}{2}}=&\; \frac{1}{\sqrt{2}} (b_{j}^{\dg}-a_{j+1}^{\dg}), \nonumber \\
\widetilde{f}^{\dg}_{j+\frac{1}{2}}=&\; \frac{1}{\sqrt{2}} (b_{j}^{\dg}+a_{j+1}^{\dg}), 
\end{align}
for $j=1,\dots,N-1$ and
\begin{align}
f^{\dg}_{\frac{1}{2}}(\alpha)=&\; \frac{1}{\sqrt{2}} (b_{N}^{\dg}-e^{-i\alpha}a_{1}^{\dg}), \nonumber \\
\widetilde{f}^{\dg}_{\frac{1}{2}}(\alpha)=&\; \frac{1}{\sqrt{2}} (b_{N}^{\dg}+ e^{-i\alpha} a_{1}^{\dg}), 
\end{align}
then the groundstate takes the form
\begin{align}
|\Psi (\alpha)\rangle_{\text{SSH}}=\prod_{j}f^{\dg}_{j+\frac{1}{2}}|0\rangle,
\end{align}
where $\prod_{j}f^{\dg}_{j+1/2}=f_{1/2}^{\dg}(\alpha)f_{3/2}^{\dg}\dots f_{N-1/2}^{\dg}$. First we compute the groundstate eigenvalue for the mirror operator $\hat{M}_{x}$ which has the following action
\begin{align}
\hat{M}_{x}: 
\begin{bmatrix}
a^{\dg} \\
b^{\dg} 
\end{bmatrix}_{j}
\mapsto
\begin{bmatrix}
b^{\dg} \\
a^{\dg} 
\end{bmatrix}_{N-j+1}.
\end{align}
therefore, it can be immediately read off that
\begin{align}
\hat{M}_{x}: \begin{bmatrix}
f^{\dg}_{j+\frac{1}{2}} \\
f^{\dg}_{\frac{1}{2}}(\alpha)
\end{bmatrix}
\mapsto
\begin{bmatrix}
-f^{\dg}_{N-j+\frac{1}{2}} \\
-e^{-i\alpha}f^{\dg}_{\frac{1}{2}}(-\alpha)
\end{bmatrix},
\end{align}
using which one can explicitly show that 
\begin{align}
{}_{\text{SSH}}\langle \Psi(\alpha) |\hat{M}_{x}|\Psi (\alpha)\rangle_{\text{SSH}}=&\; (-1)^{\frac{N}{2}(N-1)+1}\cos(\alpha).
\end{align}
This quantity by itself does not carry topological information. Furthermore when $\alpha=0$, the quantity still depends on $N$ itself and not just its parity. Instead as shown in Ref.~[\onlinecite{Shiozaki_2017}], one may define a many-body invariant as 
\begin{align}
\gamma_{\text{SSH}}:=&\; e^{\oint \mathrm{d}\alpha
{}_{\text{SSH}}\langle \Psi (\alpha) |\partial_{\alpha}|\Psi(\alpha)\rangle_{\text{SSH}}
} \nonumber \\
=&\; \frac{{}_{\text{SSH}}\langle \Psi (\pi) |\hat{M}_{x}|\Psi(\pi)\rangle_{\text{SSH}}}{{}_{\text{SSH}}\langle \Psi (0) |\hat{M}_{x}|\Psi (0)\rangle_{\text{SSH}}}=-1.
\end{align}
Further, it is known that the interaction classification of class $\ms{A}$ insulators with additional mirror reflection symmetry with $M_x^{2}=+1$ is given by the cobordism group $\Omega^{2}_{\text{pin}^{\ms{C}}}(\text{pt.})=\mathbb Z_{4}$. Therefore the above many-body invariant is not capable of detecting such a classification. In order to capture the refined interacting classification, the partial reflection operation may be used. We consider the ${U}(1)$-twisted partial mirror reflection operator $\hat{M}_{x,I}(\theta)$ which acts on the interval $I$ containing sites $j=1$ to $j=L$. The operator acts as
\begin{align}
\hat{M}_{x,I}(\theta): 
\begin{bmatrix}
a^{\dg} \\
b^{\dg} 
\end{bmatrix}_{j}
\mapsto
e^{-i\theta}
\begin{bmatrix}
b^{\dg} \\
a^{\dg} 
\end{bmatrix}_{L-j+1}.
\end{align}
While the action of $\hat{{M}}_{x,I}(\theta)$ in the bond basis takes the form
\begin{align}
\hat{M}_{\theta,I}:&\;  f^{\dg}_{\frac{1}{2}} \mapsto \left(b_
{N}^{\dg}-e^{-i\theta}b_{L}^{\dg}\right)/\sqrt{2} \nonumber \\
:&\;  f^{\dg}_{L+\frac{1}{2}} \mapsto \left(e^{-i\theta}a_
{1}^{\dg}-a_{L+1}^{\dg}\right)/\sqrt{2} \nonumber \\
:&\; f^{\dg}_{j+\frac{1}{2}}\mapsto -e^{-i\theta}f_{L-j+\frac{1}{2}}, 
\end{align}
for $j\in [1,\dots, L-1]$. For all other operators, the partial reflection acts trivially. The partial reflection eigenvalue can be computed as
\begin{widetext}
\begin{align}
{}_{\text{SSH}}\langle \Psi| \hat{M}_{\theta,I}|\Psi\rangle_{\text{SSH}}=&\; \langle 0
| f_{N-\frac{1}{2}}\dots f_{\frac{1}{2}}\hat{M}_{\theta,I}\left(f^{\dg}_{\frac{1}{2}}\underbrace{f^{\dagger}_{\frac{3}{2}}\dots f^{\dg}_{L-\frac{1}{2}}}_{L-1}
f^{\dg}_{L+\frac{1}{2}}\underbrace{f^{\dg}_{L+\frac{3}{2}}\dots f^{\dg}_{N-\frac{1}{2}}}_{N-L-1}
\right)\hat{M}_{\theta,I}^{-1}\hat{M}_{\theta,I}|0\rangle \nonumber \\
=&\; (-1)^{L-1}e^{-i(L-1)\theta}\langle 0
| f_{L+\frac{1}{2}}f_{\frac{1}{2}} f_{N-\frac{1}{2}}\dots f_{\frac{3}{2}}\underbrace{f^{\dagger}_{L-\frac{1}{2}}\dots f^{\dg}_{\frac{3}{2}}}_{L-1}
\underbrace{f^{\dg}_{L+\frac{3}{2}}\dots f^{\dg}_{N-\frac{1}{2}}}_{N-L-1}
\hat{M}_{\theta,I}\left(f^{\dg}_{\frac{1}{2}}
f^{\dg}_{L+\frac{1}{2}}
\right)\hat{M}_{\theta,I}^{-1}
\hat{M}_{\theta,I}|0\rangle \nonumber \\
=&\; (-1)^{L-1+\sum_{n=1}^{L-2}1}e^{-i(L-1)\theta}\langle 0
| f_{L+\frac{1}{2}}f_{\frac{1}{2}} 
\hat{M}_{\theta,I}\left(f^{\dg}_{\frac{1}{2}}
f^{\dg}_{L+\frac{1}{2}}
\right)\hat{M}_{\theta,I}^{-1}
\hat{M}_{\theta,I}|0\rangle \nonumber \\
=&\; \frac{1}{4}(-1)^{\frac{L}{2}(L-1)}e^{-i(L-1)\theta}
\langle 0|
f_{L+\frac{1}{2}}f_{\frac{1}{2}}\left(b_
{N}^{\dg}-e^{-i\theta}b_{L}^{\dg}\right) \left(e^{-i\theta}a_
{1}^{\dg}-a_{L+1}^{\dg}\right)
|0\rangle \nonumber \\
=&\; \frac{i}{2}e^{-iL\theta}\sin \theta (-1)^{L(L-1)/2},
\end{align}
\end{widetext}
where we have used the shorthand $M_{\theta,I}$ for $M_{x,I}(\theta)$. It can be seen that for $\theta=\pm \pi/2$ and $L$ even, the partial reflection operation produces a phase of $\pm i$ which is a topological diagnostic of $\mathbb Z_{4}=\Omega^{2}_{\text{pin}^{\ms{C}}}$. Conversely, if we consider odd $L$ i.e site-centred inversion we obtain $\text{Arg}\left({}_{\text{SSH}}\langle \Psi| \hat{M}_{\theta,0}|\Psi\rangle_{\text{SSH}}\right)\in \left\{0,\pi\right\}$ which implies a $\mathbb Z_{2}$ invariant.

\section{Cluster Perturbation Theory} \label{App:CPT}

We briefly introduce CPT and its implementation to obtain  momentum-resolved spectral functions for generalized Hubbard models on a lattice.
The basic idea behind CPT is to divide the lattice into a superlattice of clusters. The Hubbard model on each cluster is solved exactly, whereas the hoppings between sites belonging to different clusters are treated perturbatively. More details about the method and its applicability can be found in Refs.~\onlinecite{CPT-ValGros,gros1994,CPT-Sene,CPT-Sene1,CPT-Sene2,CPT-Manghi}.

We consider the general form of the Hubbard Hamiltonian:
\begin{equation} \label{eq:Hubbard_general}
    H = \sum_{\bs{r}\bs{r}'\sigma} t_{\bs{r}\sigma, \bs{r}'\sigma} c_{\bs{r}\sigma}^{\dagger} c_{\bs{r}'\sigma}
      + \sum_{\bs{r}} U n_{\bs{r}\uparrow} n_{\bs{r}\downarrow}.
\end{equation}
where $c_{\bs{r}\sigma}^{\dagger}$ ($c_{\bs{r}\sigma}$) creates (annihilates)
an electron with spin $\sigma$ at site $\bs{r}$ and 
$t_{\bs{r}\sigma, \bs{r}'\sigma}$ is the hopping amplitude  of an electron with spin $\sigma$ from site $\bs{r}'$ to  $\bs{r}$. 

The kinetic term of Eq.~\eqref{eq:Hubbard_general} can be written in a form
that shows the tiling of the lattice into clusters:
\begin{equation} \label{eq:Hubbard_general_tiled}
    H = \sum_{ij}\mathbf{c}_{i}^{\dagger}\ t^{(i,j)} \mathbf{c}_{j}
      + \sum_{\bs{r}} U n_{\bs{r}\uparrow} n_{\bs{r}\downarrow},
\end{equation}
where $i,j=1,...,L$ with $L$ the number of clusters in the crystal. $t^{(i,j)}$ is the block of the hopping matrix containing terms coupling
sites belonging to cluster $\xi_{i}$ to those of cluster $\xi_{j}$ and $\mathbf{c}_{j}$
is the column-vector of annihilation operators corresponding to sites in
cluster $\xi_{j}$. The Hamiltonian $H^{(i)}$ of a particular cluster is
obtained by choosing from Eq.~\eqref{eq:Hubbard_general_tiled} the kinetic and
interaction terms that involve only sites within the cluster $\xi_{i}$.
Mathematically, this corresponds to taking a block matrix $t^{(i,i)}$ in the diagonal of
the hopping matrix:
\begin{equation}
    H^{(i)} = \mathbf{c}_i^\dagger t^{(i,i)} \mathbf{c}_i + \sum_{\bs{r}\in\xi_{i}} U n_{\bs{r}\uparrow} n_{\bs{r}\downarrow}.
\end{equation}
The ground-state of $H^{(i)}$ is calculated with exact diagonalization~\cite{ED1, CPT-Sene2} and used to construct the cluster Green's function $G^{(i)}(\omega)$:
\begin{equation} \label{eq:Dyson_cluster}
    G^{(i)}(\omega) = \left[\omega - t^{(i,i)} - \Sigma^{(i)}(\omega)\right]^{-1},
\end{equation}
where $\Sigma^{(i)}(\omega)$ is the self-energy of $\xi_{i}$. The main approximation of CPT consists on constructing the lattice self-energy $\Sigma(\omega)$ as direct sum of cluster self-energies, i.e., as a block diagonal matrix where each block is the self-energy of a cluster: 
\begin{equation} \label{eq:CPT_approx_1}
\Sigma(\omega) = \bigoplus_{i} \Sigma^{(i)}(\omega).
\end{equation}
The Dyson equation relating the lattice Green's function $G(\omega)$ and self-energy $\Sigma(\omega)$ reads:
\begin{equation} \label{eq:Dyson_lattice}
    [G(\omega)]^{-1} = \omega - t - \Sigma(\omega),  
\end{equation}
where $t$ is the hopping matrix. Combining Eqs.~\eqref{eq:Dyson_cluster}, \eqref{eq:CPT_approx_1} and \eqref{eq:Dyson_lattice} leads to the following expression for $G(\omega)$:
\begin{equation} \label{eq:CPT_lattice_GF}
[G(\omega)]^{-1} = \bigoplus_{i} [G^{(i)}(\omega)]^{-1} - t_{\rm{inter}}.
\end{equation}
Here, $t_{\rm{inter}}$ denotes the matrix obtained by removing the blocks in the diagonal of the hopping matrix $t$, i.e. the hopping matrix including only terms that couple different clusters. Written in matrix form, Eq.\eqref{eq:CPT_lattice_GF} reads:
\begin{displaymath}
[G(\omega)]^{-1} =
\begin{pmatrix}
[G^{(1)}(\omega)]^{-1}    &    -t^{(1,2)}    &    \cdots    &    -t^{(1,L)}   \\   
-t^{(2,1)}   &    [G^{(2)}(\omega)]^{-1}    &    \cdots    &    -t^{(2,L)}   \\
\vdots    &    \vdots    &    \ddots    &    \vdots    \\
-t^{(L,1)}    &    -t^{(L,2)}    &    \cdots    &    [G^{(L)}(\omega)]^{-1}   \\
\end{pmatrix}.
\end{displaymath}
CPT inherits its name from the fact that Eq.~\eqref{eq:CPT_lattice_GF} can be derived by isolating $t_{\rm{inter}}$ in Eq.~\eqref{eq:Hubbard_general_tiled}, treating it as a perturbation to the rest and conserving only first order terms\cite{CPT-Tremblay, CPT-Sene}.

Even if position indices have not been written explicitly, the Green's function
$G(\omega)$ in Eq.~\eqref{eq:CPT_lattice_GF} is written in real-space. However,
in order to derive the momentum-resolved spectral function 
	$A(\omega, \bs{k}) = -\pi^{-1}\mathrm{Im}G(\omega+i0^{+},
	\bs{k})$ 
and the topological Hamiltonian, it is convenient to calculate its reciprocal space representation $G(\omega, \bs{k})$, by applying a \textit{periodization formula} derived below that Fourier transforms Eq.~\eqref{eq:CPT_lattice_GF} to reciprocal space.

We introduce now the concept of supercell and work out the kinematics of a lattice tiled into clusters, which will lead us to the periodization formula relating the $\bs{k}$-resolved Green function to the real-space Green function calculated by CPT. 

\begin{figure}
    \centering
    \includegraphics[width=1.0\linewidth]{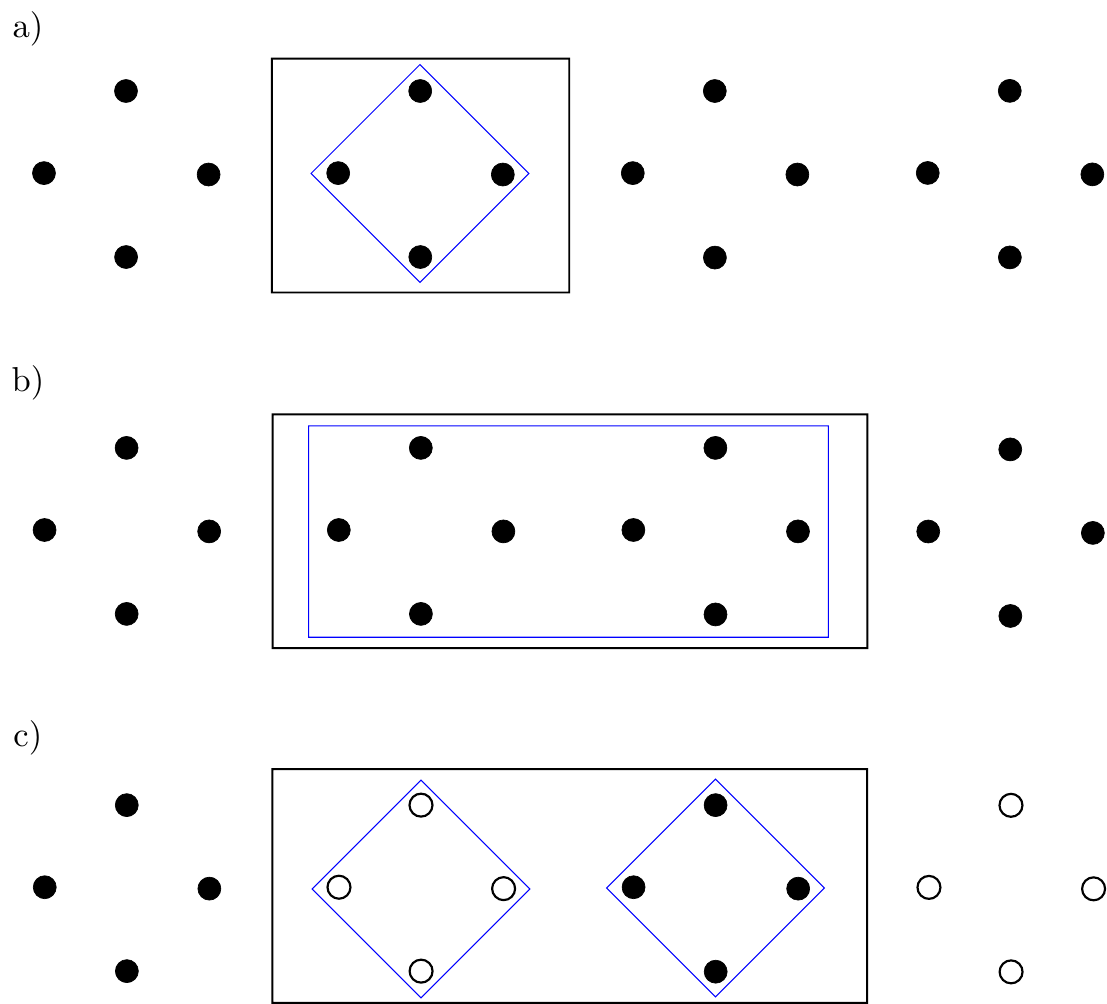}
    \caption{Different clusters and supercells of diamond-like chains. Supercells are marked with black lines, clusters with blue lines. a) Diamond chain, where a cluster contains a single diamond and a supercell a single cluster. b) Diamond chain, with a cluster containing two diamonds and a supercell containing a single cluster. c) Diamond-like chain, constructed by placing successively two different diamonds. Each cluster contains a single diamond and the smallest supercell that can be chosen contains two clusters.}
    \label{fig:tilings}
\end{figure}

A supercell is defined as a unit cell containing a group of clusters. When all clusters are of the same kind, i.e. when all $H^{(i)}$ are related by a translation of $\gamma$, a supercell containing a single cluster may be chosen (see Fig.~\ref{fig:tilings}a and Fig.~\ref{fig:tilings}b). Generally, the smallest possible supercell may contain many clusters (see Fig.~\ref{fig:tilings}c). Note that supercells form  a super-lattice $\Gamma$, which is part of the original lattice $\gamma$, so that $\Gamma \subset \gamma$. 

\begin{figure}
    \centering
    \includegraphics[width=1.0\linewidth]{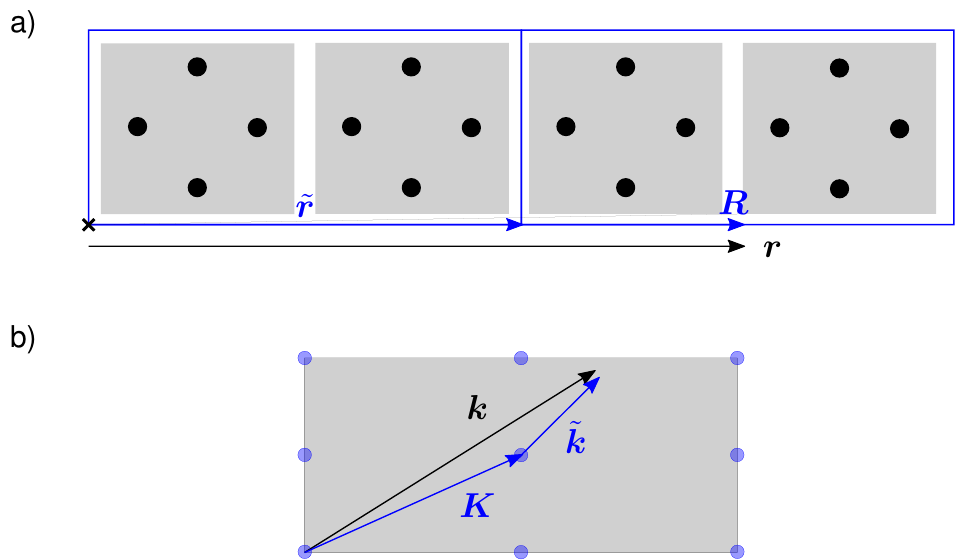}
    \caption{Illustration of the notation. (a) Example of a possible choice of the supercell (blue) in the HDC, where the unit cell is marked in grey, (b) Reciprocal structure, with the BZ in grey and sites of the superlattice denoted by blue dots.}
    \label{fig:notation_supercells}
\end{figure}

Let us consider an atom of the crystal. We denote by $\bs{r}$ the position of the unit cell it belongs to, while the corresponding supercell and unit cell within the supercell are indicated by $\tilde{\bs{r}}$ and $\bs{R}$, respectively; thus, we can write $\bs{r} = \tilde{\bs{r}} + \bs{R}$ (see Fig.\ref{fig:notation_supercells}). In reciprocal space, any vector $\bs{k}$ in the Brillouin zone of $\gamma$ can be written as $\bs{k} = \tilde{\bs{k}} + \bs{K}$, where $\tilde{\bs{k}}$ belongs to the Brillouin zone of $\Gamma$ and $\bs{K}$ to the reduced reciprocal lattice corresponding to $\Gamma$. One-body functions expressed in terms of $\bs{k}$, $\tilde{\bs{k}}$ and $\bs{K}$ are related, via the following Fourier transforms, to the descriptions that depend on $\bs{r}$, $\tilde{\bs{r}}$ and $\bs{R}$:
\begin{subequations}\label{eq:FT}
\begin{align}
& f(\bs{k}) = f(\tilde{\bs{k}}+\bs{K}) = \dfrac{1}{\sqrt{N}} \sum_{\bs{r}}
                                         e^{-i\bs{k}\cdot\bs{r}} 
                                         f(\bs{r}), \label{eq:FT1} \\ 
& f(\tilde{\bs{k}}) =  \dfrac{1}{\sqrt{N_{\Gamma}}} \sum_{\tilde{\bs{r}}}
                       e^{-i\tilde{\bs{k}}\cdot\tilde{\bs{r}}} 
                       f(\tilde{\bs{r}}), \label{eq:FT2}\\
& f(\bs{K}) =  \dfrac{\sqrt{N_{\Gamma}}}{\sqrt{N}} \sum_{\bs{R}}
               e^{-i\bs{K}\cdot\bs{R}} 
               f(\bs{R}). \label{eq:FT3}
\end{align}
\end{subequations}
where $N_{\Gamma}$ is the number of supercells in the lattice. We can define two reciprocal space representations: On the one hand, the $\bs{k}$-representation, based on the transformation of Eq.~\eqref{eq:FT1}. On the other hand, the $(\tilde{\bs{k}}, \bs{K})$-representation is obtained by consecutive application of Eq.~\eqref{eq:FT2} and Eq.~\eqref{eq:FT3}. Each of these representations relates the real and reciprocal space representations of the annihilation operators in the following form:
\begin{subequations}
\begin{align}
& c({\bs{k}}) = \dfrac{1}{\sqrt{N}} \sum_{r} e^{-i\bs{k}\cdot\bs{r}} c(\bs{r}), \label{eq:k-repr} \\
& c_{\bs{K}}(\tilde{\bs{k}}) = \dfrac{1}{\sqrt{N}} \sum_{\bs{R}} \sum_{\bs{r}} e^{-i\tilde{\bs{k}}\cdot\tilde{\bs{r}}}e^{-i\bs{K}\cdot\bs{R}} c(\tilde{\bs{r}}+\bs{R}). \label{eq:kK-repr}
\end{align}
\end{subequations}
Here, Eq.~\eqref{eq:k-repr} corresponds to the $\bs{k}$-representation and Eq.~\eqref{eq:kK-repr} to the $(\tilde{\bs{k}}, \bs{K})$-representation. The matrix $\Delta$ relating both representations, $c(\bs{k}) = \Delta_{\bs{k},\tilde{\bs{k}}'\bs{K}'} c_{\bs{K}'}(\tilde{\bs{k}}')$, is the following:
\begin{equation} \label{eq:CPT_delta}
\begin{split}
\Delta_{\bs{k},\tilde{\bs{k}}'\bs{K}'} &= \dfrac{1}{N} \sum_{\tilde{\bs{r}}\bs{R}}
                                        e^{-i\bs{k}\cdot(\tilde{\bs{r}}+\bs{R})} e^{i\tilde{\bs{k}}'\cdot\tilde{\bs{r}}} e^{i\bs{K}'\cdot\bs{R}} \\
                                     &= \delta_{\tilde{\bs{k}}\tilde{\bs{k}}'}
                                        \dfrac{N_{\Gamma}}{N} \sum_{\bs{R}} e^{-i(\tilde{\bs{k}}+\bs{K}-\bs{K}')\cdot\bs{R}}.
\end{split}
\end{equation}
where $\bs{k} = \tilde{\bs{k}} + \bs{K}$. Note that $\Delta$ is not diagonal for all $\bs{k}$ and $(\tilde{\bs{k}}', \bs{K}')$, which means that the representations are not equivalent. 

In addition, we can set a $(\tilde{\bs{k}},\bs{R})$-representation, which lies midway between both, $\bs{k}$ and $(\tilde{\bs{k}}, \bs{K})$-representations:
\begin{equation}
c_{\bs{R}}(\tilde{\bs{k}}) = \dfrac{1}{\sqrt{N_{\Gamma}}} \sum_{\tilde{\bs{r}}}
                        e^{-i\tilde{\bs{k}}\cdot\tilde{\bs{r}}} c(\tilde{\bs{r}}+\bs{R}).
\end{equation}
Generally, treating inter and intra-cluster hoppings differently breaks the invariance under translations of $\gamma$, thus $G(\omega)$ in Eq.~\eqref{eq:CPT_lattice_GF} is not diagonal in the $\bs{k}$-representation. Nevertheless, since invariance under translations of $\Gamma$ is preserved, it is diagonal in $\tilde{\bs{k}}$-indices. Therefore, it is convenient to express $G(\omega)$ in the $(\tilde{\bs{k}}, \bs{R})$ representation:
\begin{equation}
    [G(\omega, \tilde{\bs{k}})]^{-1} = \bigoplus_{i} [G^{(i)}(\omega)]^{-1} - t_{\rm{inter}}(\tilde{\bs{k}}),
\end{equation}
where $\bs{R}$ and $\bs{R}'$ indices have been omitted. In order to achieve the $\bs{k}$-representation of the Green function, we first write it in the $(\tilde{\bs{k}}, \bs{K})$-representation:
\begin{equation}
G_{\bs{K}\bs{K}'}(\omega, \tilde{\bs{k}}) = \dfrac{N_{\Gamma}}{N}\sum_{\bs{R}\bs{R}'}
                                    e^{i\bs{K}'\cdot\bs{R}'} e^{-i\bs{K}\cdot\bs{R}}
                                    G_{\bs{R}\bs{R}'}(\omega, \tilde{\bs{k}}).
\end{equation}
Applying the matrix $\Delta$ of Eq.~\eqref{eq:CPT_delta} to $G_{\bs{K}\bs{K}'}(\omega, \tilde{\bs{k}})$ leads to the following expression of the Green function in the $\bs{k}$-representation:
\begin{equation} \label{eq:k-repr-non-diag}
    G(\omega, \bs{k},\bs{k}') = \dfrac{N_{\Gamma}}{N}\sum_{\bs{R}\bs{R}'}
                                e^{i\bs{k}'\cdot\bs{R}'} e^{-i\bs{k}\cdot\bs{R}}
                                G_{\bs{R}\bs{R}'}(\omega, \tilde{\bs{k}}),
\end{equation}
where $\bs{k}=\tilde{\bs{k}}+\bs{K}$ and $\bs{k}'=\tilde{\bs{k}}+\bs{K}'$. Note that we can make the substitution $\tilde{\bs{k}} \rightarrow \bs{k} = \tilde{\bs{k}} + \bs{K}$ in the Green function $G_{\bs{R}\bs{R}'}(\omega, \tilde{\bs{k}})$ on the right-side, since $\tilde{\bs{k}}$ is a vector defined up to a vector $\bs{K}$ belonging to the reciprocal lattice of $\Gamma$.  Note also that, since $\bs{K}$ and $\bs{K}'$ may not be identical, the Green function in Eq.~\eqref{eq:k-repr-non-diag} is not diagonal in $\bs{k}=\tilde{\bs{k}}+\bs{K}$ and $\bs{k}'=\tilde{\bs{k}}+\bs{K}'$. For an element in the diagonal, the expression takes the following form, often called \textit{periodization formula}:
\begin{equation}
G(\omega, \bs{k}) = \dfrac{N_{\Gamma}}{N} \sum_{\bs{R}\bs{R}'}
                    e^{i\bs{k}\cdot(\bs{R}'-\bs{R})}
                    G_{\bs{R}\bs{R}'}(\omega, \bs{k}).
\end{equation}
The periodization formula contains all the information needed to compute the density of states $\rho(\omega)$, as this only involves diagonal elements of the $\bs{k}$-representation of the Green function,
\begin{equation} \label{eq:dos}
\rho(\omega) = -\dfrac{1}{\pi} \sum_{\bs{k}} \mathrm{Im}G(\omega+i0^{+}, \bs{k}),
\end{equation}
Moreover, the $\bs{k}$-resolved spectral function $A(\omega, \bs{k})$ also involves only elements on the diagonal.

Considering that the normalization of $\rho(\omega)$ reads $\int_{-\infty}^{\infty} d\omega\ \rho(\omega)=1$, the chemical potential $\mu$ is computed from $\rho(\omega)$  based on the following formula:
\begin{equation} \label{eq:def_mu}
\int_{-\infty}^{\mu} d\omega\  \rho(\omega) = 1/2.
\end{equation}
Eq.~\eqref{eq:def_mu} can also be considered the equation defining $\mu$.

\bibliography{main}

\begin{thebibliography}{74}%
\makeatletter
\providecommand \@ifxundefined [1]{%
 \@ifx{#1\undefined}
}%
\providecommand \@ifnum [1]{%
 \ifnum #1\expandafter \@firstoftwo
 \else \expandafter \@secondoftwo
 \fi
}%
\providecommand \@ifx [1]{%
 \ifx #1\expandafter \@firstoftwo
 \else \expandafter \@secondoftwo
 \fi
}%
\providecommand \natexlab [1]{#1}%
\providecommand \enquote  [1]{``#1''}%
\providecommand \bibnamefont  [1]{#1}%
\providecommand \bibfnamefont [1]{#1}%
\providecommand \citenamefont [1]{#1}%
\providecommand \href@noop [0]{\@secondoftwo}%
\providecommand \href [0]{\begingroup \@sanitize@url \@href}%
\providecommand \@href[1]{\@@startlink{#1}\@@href}%
\providecommand \@@href[1]{\endgroup#1\@@endlink}%
\providecommand \@sanitize@url [0]{\catcode `\\12\catcode `\$12\catcode
  `\&12\catcode `\#12\catcode `\^12\catcode `\_12\catcode `\%12\relax}%
\providecommand \@@startlink[1]{}%
\providecommand \@@endlink[0]{}%
\providecommand \url  [0]{\begingroup\@sanitize@url \@url }%
\providecommand \@url [1]{\endgroup\@href {#1}{\urlprefix }}%
\providecommand \urlprefix  [0]{URL }%
\providecommand \Eprint [0]{\href }%
\providecommand \doibase [0]{http://dx.doi.org/}%
\providecommand \selectlanguage [0]{\@gobble}%
\providecommand \bibinfo  [0]{\@secondoftwo}%
\providecommand \bibfield  [0]{\@secondoftwo}%
\providecommand \translation [1]{[#1]}%
\providecommand \BibitemOpen [0]{}%
\providecommand \bibitemStop [0]{}%
\providecommand \bibitemNoStop [0]{.\EOS\space}%
\providecommand \EOS [0]{\spacefactor3000\relax}%
\providecommand \BibitemShut  [1]{\csname bibitem#1\endcsname}%
\let\auto@bib@innerbib\@empty
\bibitem [{\citenamefont {Hasan}\ and\ \citenamefont
  {Kane}(2010)}]{CharlieZahidReview}%
  \BibitemOpen
  \bibfield  {author} {\bibinfo {author} {\bibfnamefont {M.~Z.}\ \bibnamefont
  {Hasan}}\ and\ \bibinfo {author} {\bibfnamefont {C.~L.}\ \bibnamefont
  {Kane}},\ }\href {\doibase 10.1103/RevModPhys.82.3045} {\bibfield  {journal}
  {\bibinfo  {journal} {Rev. Mod. Phys.}\ }\textbf {\bibinfo {volume} {82}},\
  \bibinfo {pages} {3045} (\bibinfo {year} {2010})}\BibitemShut {NoStop}%
\bibitem [{\citenamefont {Kane}\ and\ \citenamefont
  {Mele}(2005)}]{PhysRevLett.95.146802}%
  \BibitemOpen
  \bibfield  {author} {\bibinfo {author} {\bibfnamefont {C.~L.}\ \bibnamefont
  {Kane}}\ and\ \bibinfo {author} {\bibfnamefont {E.~J.}\ \bibnamefont
  {Mele}},\ }\href {\doibase 10.1103/PhysRevLett.95.146802} {\bibfield
  {journal} {\bibinfo  {journal} {Phys. Rev. Lett.}\ }\textbf {\bibinfo
  {volume} {95}},\ \bibinfo {pages} {146802} (\bibinfo {year}
  {2005})}\BibitemShut {NoStop}%
\bibitem [{\citenamefont {Bernevig}\ \emph {et~al.}(2006)\citenamefont
  {Bernevig}, \citenamefont {Hughes},\ and\ \citenamefont
  {Zhang}}]{Bernevig1757}%
  \BibitemOpen
  \bibfield  {author} {\bibinfo {author} {\bibfnamefont {B.~A.}\ \bibnamefont
  {Bernevig}}, \bibinfo {author} {\bibfnamefont {T.~L.}\ \bibnamefont
  {Hughes}}, \ and\ \bibinfo {author} {\bibfnamefont {S.-C.}\ \bibnamefont
  {Zhang}},\ }\href {\doibase 10.1126/science.1133734} {\bibfield  {journal}
  {\bibinfo  {journal} {Science}\ }\textbf {\bibinfo {volume} {314}},\ \bibinfo
  {pages} {1757} (\bibinfo {year} {2006})},\ \Eprint
  {http://arxiv.org/abs/https://science.sciencemag.org/content/314/5806/1757.full.pdf}
  {https://science.sciencemag.org/content/314/5806/1757.full.pdf} \BibitemShut
  {NoStop}%
\bibitem [{\citenamefont {von Klitzing}(1986)}]{RevModPhys.58.519}%
  \BibitemOpen
  \bibfield  {author} {\bibinfo {author} {\bibfnamefont {K.}~\bibnamefont {von
  Klitzing}},\ }\href {\doibase 10.1103/RevModPhys.58.519} {\bibfield
  {journal} {\bibinfo  {journal} {Rev. Mod. Phys.}\ }\textbf {\bibinfo {volume}
  {58}},\ \bibinfo {pages} {519} (\bibinfo {year} {1986})}\BibitemShut
  {NoStop}%
\bibitem [{\citenamefont {Tsui}\ \emph {et~al.}(1982)\citenamefont {Tsui},
  \citenamefont {Stormer},\ and\ \citenamefont
  {Gossard}}]{PhysRevLett.48.1559}%
  \BibitemOpen
  \bibfield  {author} {\bibinfo {author} {\bibfnamefont {D.~C.}\ \bibnamefont
  {Tsui}}, \bibinfo {author} {\bibfnamefont {H.~L.}\ \bibnamefont {Stormer}}, \
  and\ \bibinfo {author} {\bibfnamefont {A.~C.}\ \bibnamefont {Gossard}},\
  }\href {\doibase 10.1103/PhysRevLett.48.1559} {\bibfield  {journal} {\bibinfo
   {journal} {Phys. Rev. Lett.}\ }\textbf {\bibinfo {volume} {48}},\ \bibinfo
  {pages} {1559} (\bibinfo {year} {1982})}\BibitemShut {NoStop}%
\bibitem [{\citenamefont {Tsui}(1999)}]{RevModPhys.71.891}%
  \BibitemOpen
  \bibfield  {author} {\bibinfo {author} {\bibfnamefont {D.~C.}\ \bibnamefont
  {Tsui}},\ }\href {\doibase 10.1103/RevModPhys.71.891} {\bibfield  {journal}
  {\bibinfo  {journal} {Rev. Mod. Phys.}\ }\textbf {\bibinfo {volume} {71}},\
  \bibinfo {pages} {891} (\bibinfo {year} {1999})}\BibitemShut {NoStop}%
\bibitem [{\citenamefont {Bradlyn}\ \emph {et~al.}(2017)\citenamefont
  {Bradlyn}, \citenamefont {Elcoro}, \citenamefont {Cano}, \citenamefont
  {Vergniory}, \citenamefont {Wang}, \citenamefont {Felser}, \citenamefont
  {Aroyo},\ and\ \citenamefont {Bernevig}}]{Bradlyn_TQC}%
  \BibitemOpen
  \bibfield  {author} {\bibinfo {author} {\bibfnamefont {B.}~\bibnamefont
  {Bradlyn}}, \bibinfo {author} {\bibfnamefont {L.}~\bibnamefont {Elcoro}},
  \bibinfo {author} {\bibfnamefont {J.}~\bibnamefont {Cano}}, \bibinfo {author}
  {\bibfnamefont {M.~G.}\ \bibnamefont {Vergniory}}, \bibinfo {author}
  {\bibfnamefont {Z.}~\bibnamefont {Wang}}, \bibinfo {author} {\bibfnamefont
  {C.}~\bibnamefont {Felser}}, \bibinfo {author} {\bibfnamefont {M.~I.}\
  \bibnamefont {Aroyo}}, \ and\ \bibinfo {author} {\bibfnamefont {B.~A.}\
  \bibnamefont {Bernevig}},\ }\href {\doibase 10.1038/nature23268} {\bibfield
  {journal} {\bibinfo  {journal} {Nature}\ }\textbf {\bibinfo {volume} {547}},\
  \bibinfo {pages} {298} (\bibinfo {year} {2017})}\BibitemShut {NoStop}%
\bibitem [{\citenamefont {Song}\ \emph
  {et~al.}(2018{\natexlab{a}})\citenamefont {Song}, \citenamefont {Zhang},\
  and\ \citenamefont {Fang}}]{ZhidaSemimetals}%
  \BibitemOpen
  \bibfield  {author} {\bibinfo {author} {\bibfnamefont {Z.}~\bibnamefont
  {Song}}, \bibinfo {author} {\bibfnamefont {T.}~\bibnamefont {Zhang}}, \ and\
  \bibinfo {author} {\bibfnamefont {C.}~\bibnamefont {Fang}},\ }\href {\doibase
  10.1103/PhysRevX.8.031069} {\bibfield  {journal} {\bibinfo  {journal} {Phys.
  Rev. X}\ }\textbf {\bibinfo {volume} {8}},\ \bibinfo {pages} {031069}
  (\bibinfo {year} {2018}{\natexlab{a}})}\BibitemShut {NoStop}%
\bibitem [{\citenamefont {Po}\ \emph {et~al.}(2017{\natexlab{a}})\citenamefont
  {Po}, \citenamefont {Vishwanath},\ and\ \citenamefont
  {Watanabe}}]{AshvinIndicators}%
  \BibitemOpen
  \bibfield  {author} {\bibinfo {author} {\bibfnamefont {H.~C.}\ \bibnamefont
  {Po}}, \bibinfo {author} {\bibfnamefont {A.}~\bibnamefont {Vishwanath}}, \
  and\ \bibinfo {author} {\bibfnamefont {H.}~\bibnamefont {Watanabe}},\ }\href
  {\doibase 10.1038/s41467-017-00133-2} {\bibfield  {journal} {\bibinfo
  {journal} {Nature Communications}\ }\textbf {\bibinfo {volume} {8}},\
  \bibinfo {pages} {50} (\bibinfo {year} {2017}{\natexlab{a}})}\BibitemShut
  {NoStop}%
\bibitem [{\citenamefont {Song}\ \emph
  {et~al.}(2018{\natexlab{b}})\citenamefont {Song}, \citenamefont {Zhang},
  \citenamefont {Fang},\ and\ \citenamefont {Fang}}]{ChenTCI}%
  \BibitemOpen
  \bibfield  {author} {\bibinfo {author} {\bibfnamefont {Z.}~\bibnamefont
  {Song}}, \bibinfo {author} {\bibfnamefont {T.}~\bibnamefont {Zhang}},
  \bibinfo {author} {\bibfnamefont {Z.}~\bibnamefont {Fang}}, \ and\ \bibinfo
  {author} {\bibfnamefont {C.}~\bibnamefont {Fang}},\ }\href {\doibase
  10.1038/s41467-018-06010-w} {\bibfield  {journal} {\bibinfo  {journal}
  {Nature Communications}\ }\textbf {\bibinfo {volume} {9}},\ \bibinfo {pages}
  {3530} (\bibinfo {year} {2018}{\natexlab{b}})}\BibitemShut {NoStop}%
\bibitem [{\citenamefont {Rachel}(2018)}]{Rachel_2018}%
  \BibitemOpen
  \bibfield  {author} {\bibinfo {author} {\bibfnamefont {S.}~\bibnamefont
  {Rachel}},\ }\href {\doibase 10.1088/1361-6633/aad6a6} {\bibfield  {journal}
  {\bibinfo  {journal} {Reports on Progress in Physics}\ }\textbf {\bibinfo
  {volume} {81}},\ \bibinfo {pages} {116501} (\bibinfo {year}
  {2018})}\BibitemShut {NoStop}%
\bibitem [{\citenamefont {Gu}\ and\ \citenamefont
  {Wen}(2009)}]{PhysRevB.80.155131}%
  \BibitemOpen
  \bibfield  {author} {\bibinfo {author} {\bibfnamefont {Z.-C.}\ \bibnamefont
  {Gu}}\ and\ \bibinfo {author} {\bibfnamefont {X.-G.}\ \bibnamefont {Wen}},\
  }\href {\doibase 10.1103/PhysRevB.80.155131} {\bibfield  {journal} {\bibinfo
  {journal} {Phys. Rev. B}\ }\textbf {\bibinfo {volume} {80}},\ \bibinfo
  {pages} {155131} (\bibinfo {year} {2009})}\BibitemShut {NoStop}%
\bibitem [{\citenamefont {Pollmann}\ \emph {et~al.}(2010)\citenamefont
  {Pollmann}, \citenamefont {Turner}, \citenamefont {Berg},\ and\ \citenamefont
  {Oshikawa}}]{PhysRevB.81.064439}%
  \BibitemOpen
  \bibfield  {author} {\bibinfo {author} {\bibfnamefont {F.}~\bibnamefont
  {Pollmann}}, \bibinfo {author} {\bibfnamefont {A.~M.}\ \bibnamefont
  {Turner}}, \bibinfo {author} {\bibfnamefont {E.}~\bibnamefont {Berg}}, \ and\
  \bibinfo {author} {\bibfnamefont {M.}~\bibnamefont {Oshikawa}},\ }\href
  {\doibase 10.1103/PhysRevB.81.064439} {\bibfield  {journal} {\bibinfo
  {journal} {Phys. Rev. B}\ }\textbf {\bibinfo {volume} {81}},\ \bibinfo
  {pages} {064439} (\bibinfo {year} {2010})}\BibitemShut {NoStop}%
\bibitem [{\citenamefont {Fidkowski}\ and\ \citenamefont
  {Kitaev}(2011)}]{PhysRevB.83.075103}%
  \BibitemOpen
  \bibfield  {author} {\bibinfo {author} {\bibfnamefont {L.}~\bibnamefont
  {Fidkowski}}\ and\ \bibinfo {author} {\bibfnamefont {A.}~\bibnamefont
  {Kitaev}},\ }\href {\doibase 10.1103/PhysRevB.83.075103} {\bibfield
  {journal} {\bibinfo  {journal} {Phys. Rev. B}\ }\textbf {\bibinfo {volume}
  {83}},\ \bibinfo {pages} {075103} (\bibinfo {year} {2011})}\BibitemShut
  {NoStop}%
\bibitem [{\citenamefont {Turner}\ \emph {et~al.}(2011)\citenamefont {Turner},
  \citenamefont {Pollmann},\ and\ \citenamefont {Berg}}]{PhysRevB.83.075102}%
  \BibitemOpen
  \bibfield  {author} {\bibinfo {author} {\bibfnamefont {A.~M.}\ \bibnamefont
  {Turner}}, \bibinfo {author} {\bibfnamefont {F.}~\bibnamefont {Pollmann}}, \
  and\ \bibinfo {author} {\bibfnamefont {E.}~\bibnamefont {Berg}},\ }\href
  {\doibase 10.1103/PhysRevB.83.075102} {\bibfield  {journal} {\bibinfo
  {journal} {Phys. Rev. B}\ }\textbf {\bibinfo {volume} {83}},\ \bibinfo
  {pages} {075102} (\bibinfo {year} {2011})}\BibitemShut {NoStop}%
\bibitem [{\citenamefont {Chen}\ \emph {et~al.}(2011)\citenamefont {Chen},
  \citenamefont {Gu},\ and\ \citenamefont {Wen}}]{PhysRevB.83.035107}%
  \BibitemOpen
  \bibfield  {author} {\bibinfo {author} {\bibfnamefont {X.}~\bibnamefont
  {Chen}}, \bibinfo {author} {\bibfnamefont {Z.-C.}\ \bibnamefont {Gu}}, \ and\
  \bibinfo {author} {\bibfnamefont {X.-G.}\ \bibnamefont {Wen}},\ }\href
  {\doibase 10.1103/PhysRevB.83.035107} {\bibfield  {journal} {\bibinfo
  {journal} {Phys. Rev. B}\ }\textbf {\bibinfo {volume} {83}},\ \bibinfo
  {pages} {035107} (\bibinfo {year} {2011})}\BibitemShut {NoStop}%
\bibitem [{\citenamefont {Senthil}(2015)}]{shentil-rev}%
  \BibitemOpen
  \bibfield  {author} {\bibinfo {author} {\bibfnamefont {T.}~\bibnamefont
  {Senthil}},\ }\href {\doibase 10.1146/annurev-conmatphys-031214-014740}
  {\bibfield  {journal} {\bibinfo  {journal} {Annual Review of Condensed Matter
  Physics}\ }\textbf {\bibinfo {volume} {6}},\ \bibinfo {pages} {299} (\bibinfo
  {year} {2015})},\ \Eprint
  {http://arxiv.org/abs/https://doi.org/10.1146/annurev-conmatphys-031214-014740}
  {https://doi.org/10.1146/annurev-conmatphys-031214-014740} \BibitemShut
  {NoStop}%
\bibitem [{\citenamefont {Song}\ \emph {et~al.}(2017)\citenamefont {Song},
  \citenamefont {Huang}, \citenamefont {Fu},\ and\ \citenamefont
  {Hermele}}]{PhysRevX.7.011020}%
  \BibitemOpen
  \bibfield  {author} {\bibinfo {author} {\bibfnamefont {H.}~\bibnamefont
  {Song}}, \bibinfo {author} {\bibfnamefont {S.-J.}\ \bibnamefont {Huang}},
  \bibinfo {author} {\bibfnamefont {L.}~\bibnamefont {Fu}}, \ and\ \bibinfo
  {author} {\bibfnamefont {M.}~\bibnamefont {Hermele}},\ }\href {\doibase
  10.1103/PhysRevX.7.011020} {\bibfield  {journal} {\bibinfo  {journal} {Phys.
  Rev. X}\ }\textbf {\bibinfo {volume} {7}},\ \bibinfo {pages} {011020}
  (\bibinfo {year} {2017})}\BibitemShut {NoStop}%
\bibitem [{\citenamefont {Fu}\ \emph {et~al.}(2007)\citenamefont {Fu},
  \citenamefont {Kane},\ and\ \citenamefont {Mele}}]{PhysRevLett.98.106803}%
  \BibitemOpen
  \bibfield  {author} {\bibinfo {author} {\bibfnamefont {L.}~\bibnamefont
  {Fu}}, \bibinfo {author} {\bibfnamefont {C.~L.}\ \bibnamefont {Kane}}, \ and\
  \bibinfo {author} {\bibfnamefont {E.~J.}\ \bibnamefont {Mele}},\ }\href
  {\doibase 10.1103/PhysRevLett.98.106803} {\bibfield  {journal} {\bibinfo
  {journal} {Phys. Rev. Lett.}\ }\textbf {\bibinfo {volume} {98}},\ \bibinfo
  {pages} {106803} (\bibinfo {year} {2007})}\BibitemShut {NoStop}%
\bibitem [{\citenamefont {Altland}\ and\ \citenamefont
  {Zirnbauer}(1997)}]{PhysRevB.55.1142}%
  \BibitemOpen
  \bibfield  {author} {\bibinfo {author} {\bibfnamefont {A.}~\bibnamefont
  {Altland}}\ and\ \bibinfo {author} {\bibfnamefont {M.~R.}\ \bibnamefont
  {Zirnbauer}},\ }\href {\doibase 10.1103/PhysRevB.55.1142} {\bibfield
  {journal} {\bibinfo  {journal} {Phys. Rev. B}\ }\textbf {\bibinfo {volume}
  {55}},\ \bibinfo {pages} {1142} (\bibinfo {year} {1997})}\BibitemShut
  {NoStop}%
\bibitem [{\citenamefont {Chiu}\ \emph {et~al.}(2016)\citenamefont {Chiu},
  \citenamefont {Teo}, \citenamefont {Schnyder},\ and\ \citenamefont
  {Ryu}}]{RevModPhys.88.035005}%
  \BibitemOpen
  \bibfield  {author} {\bibinfo {author} {\bibfnamefont {C.-K.}\ \bibnamefont
  {Chiu}}, \bibinfo {author} {\bibfnamefont {J.~C.~Y.}\ \bibnamefont {Teo}},
  \bibinfo {author} {\bibfnamefont {A.~P.}\ \bibnamefont {Schnyder}}, \ and\
  \bibinfo {author} {\bibfnamefont {S.}~\bibnamefont {Ryu}},\ }\href {\doibase
  10.1103/RevModPhys.88.035005} {\bibfield  {journal} {\bibinfo  {journal}
  {Rev. Mod. Phys.}\ }\textbf {\bibinfo {volume} {88}},\ \bibinfo {pages}
  {035005} (\bibinfo {year} {2016})}\BibitemShut {NoStop}%
\bibitem [{\citenamefont {Fu}(2011)}]{PhysRevLett.106.106802}%
  \BibitemOpen
  \bibfield  {author} {\bibinfo {author} {\bibfnamefont {L.}~\bibnamefont
  {Fu}},\ }\href {\doibase 10.1103/PhysRevLett.106.106802} {\bibfield
  {journal} {\bibinfo  {journal} {Phys. Rev. Lett.}\ }\textbf {\bibinfo
  {volume} {106}},\ \bibinfo {pages} {106802} (\bibinfo {year}
  {2011})}\BibitemShut {NoStop}%
\bibitem [{\citenamefont {Soluyanov}\ and\ \citenamefont
  {Vanderbilt}(2012)}]{PhysRevB.85.115415}%
  \BibitemOpen
  \bibfield  {author} {\bibinfo {author} {\bibfnamefont {A.~A.}\ \bibnamefont
  {Soluyanov}}\ and\ \bibinfo {author} {\bibfnamefont {D.}~\bibnamefont
  {Vanderbilt}},\ }\href {\doibase 10.1103/PhysRevB.85.115415} {\bibfield
  {journal} {\bibinfo  {journal} {Phys. Rev. B}\ }\textbf {\bibinfo {volume}
  {85}},\ \bibinfo {pages} {115415} (\bibinfo {year} {2012})}\BibitemShut
  {NoStop}%
\bibitem [{\citenamefont {Vergniory}\ \emph {et~al.}(2019)\citenamefont
  {Vergniory}, \citenamefont {Elcoro}, \citenamefont {Felser}, \citenamefont
  {Regnault}, \citenamefont {Bernevig},\ and\ \citenamefont {Wang}}]{vergnio}%
  \BibitemOpen
  \bibfield  {author} {\bibinfo {author} {\bibfnamefont {M.~G.}\ \bibnamefont
  {Vergniory}}, \bibinfo {author} {\bibfnamefont {L.}~\bibnamefont {Elcoro}},
  \bibinfo {author} {\bibfnamefont {C.}~\bibnamefont {Felser}}, \bibinfo
  {author} {\bibfnamefont {N.}~\bibnamefont {Regnault}}, \bibinfo {author}
  {\bibfnamefont {B.~A.}\ \bibnamefont {Bernevig}}, \ and\ \bibinfo {author}
  {\bibfnamefont {Z.}~\bibnamefont {Wang}},\ }\href {\doibase
  10.1038/s41586-019-0954-4} {\bibfield  {journal} {\bibinfo  {journal}
  {Nature}\ }\textbf {\bibinfo {volume} {566}},\ \bibinfo {pages} {480}
  (\bibinfo {year} {2019})}\BibitemShut {NoStop}%
\bibitem [{\citenamefont {Zak}(1980)}]{PhysRevLett.45.1025}%
  \BibitemOpen
  \bibfield  {author} {\bibinfo {author} {\bibfnamefont {J.}~\bibnamefont
  {Zak}},\ }\href {\doibase 10.1103/PhysRevLett.45.1025} {\bibfield  {journal}
  {\bibinfo  {journal} {Phys. Rev. Lett.}\ }\textbf {\bibinfo {volume} {45}},\
  \bibinfo {pages} {1025} (\bibinfo {year} {1980})}\BibitemShut {NoStop}%
\bibitem [{\citenamefont {Michel}\ and\ \citenamefont
  {Zak}(1999)}]{Michel1999}%
  \BibitemOpen
  \bibfield  {author} {\bibinfo {author} {\bibfnamefont {L.}~\bibnamefont
  {Michel}}\ and\ \bibinfo {author} {\bibfnamefont {J.}~\bibnamefont {Zak}},\
  }\href@noop {} {\bibfield  {journal} {\bibinfo  {journal} {Phys. Rev. B}\
  }\textbf {\bibinfo {volume} {59}},\ \bibinfo {pages} {5998} (\bibinfo {year}
  {1999})}\BibitemShut {NoStop}%
\bibitem [{\citenamefont {Michel}\ and\ \citenamefont
  {Zak}(2001)}]{Michel2001}%
  \BibitemOpen
  \bibfield  {author} {\bibinfo {author} {\bibfnamefont {L.}~\bibnamefont
  {Michel}}\ and\ \bibinfo {author} {\bibfnamefont {J.}~\bibnamefont {Zak}},\
  }\href@noop {} {\bibfield  {journal} {\bibinfo  {journal} {Phys. Rep.}\
  }\textbf {\bibinfo {volume} {341}},\ \bibinfo {pages} {377} (\bibinfo {year}
  {2001})}\BibitemShut {NoStop}%
\bibitem [{\citenamefont {Yao}\ and\ \citenamefont
  {Kivelson}(2010)}]{Kivelson10}%
  \BibitemOpen
  \bibfield  {author} {\bibinfo {author} {\bibfnamefont {H.}~\bibnamefont
  {Yao}}\ and\ \bibinfo {author} {\bibfnamefont {S.~A.}\ \bibnamefont
  {Kivelson}},\ }\href {\doibase 10.1103/PhysRevLett.105.166402} {\bibfield
  {journal} {\bibinfo  {journal} {Phys. Rev. Lett.}\ }\textbf {\bibinfo
  {volume} {105}},\ \bibinfo {pages} {166402} (\bibinfo {year}
  {2010})}\BibitemShut {NoStop}%
\bibitem [{\citenamefont {Fuji}\ \emph {et~al.}(2015)\citenamefont {Fuji},
  \citenamefont {Pollmann},\ and\ \citenamefont {Oshikawa}}]{fuji2015}%
  \BibitemOpen
  \bibfield  {author} {\bibinfo {author} {\bibfnamefont {Y.}~\bibnamefont
  {Fuji}}, \bibinfo {author} {\bibfnamefont {F.}~\bibnamefont {Pollmann}}, \
  and\ \bibinfo {author} {\bibfnamefont {M.}~\bibnamefont {Oshikawa}},\
  }\href@noop {} {\bibfield  {journal} {\bibinfo  {journal} {Physical Review
  Letters}\ }\textbf {\bibinfo {volume} {114}},\ \bibinfo {pages} {177204}
  (\bibinfo {year} {2015})}\BibitemShut {NoStop}%
\bibitem [{\citenamefont {Po}\ \emph {et~al.}(2017{\natexlab{b}})\citenamefont
  {Po}, \citenamefont {Vishwanath},\ and\ \citenamefont
  {Watanabe}}]{Ashwin_sym}%
  \BibitemOpen
  \bibfield  {author} {\bibinfo {author} {\bibfnamefont {H.~C.}\ \bibnamefont
  {Po}}, \bibinfo {author} {\bibfnamefont {A.}~\bibnamefont {Vishwanath}}, \
  and\ \bibinfo {author} {\bibfnamefont {H.}~\bibnamefont {Watanabe}},\ }\href
  {\doibase 10.1038/s41467-017-00133-2} {\bibfield  {journal} {\bibinfo
  {journal} {Nature Communications}\ }\textbf {\bibinfo {volume} {8}},\
  \bibinfo {pages} {50} (\bibinfo {year} {2017}{\natexlab{b}})}\BibitemShut
  {NoStop}%
\bibitem [{\citenamefont {Gurarie}(2011)}]{gurarieG}%
  \BibitemOpen
  \bibfield  {author} {\bibinfo {author} {\bibfnamefont {V.}~\bibnamefont
  {Gurarie}},\ }\href {\doibase 10.1103/PhysRevB.83.085426} {\bibfield
  {journal} {\bibinfo  {journal} {Phys. Rev. B}\ }\textbf {\bibinfo {volume}
  {83}},\ \bibinfo {pages} {085426} (\bibinfo {year} {2011})}\BibitemShut
  {NoStop}%
\bibitem [{\citenamefont {Wang}\ and\ \citenamefont {Zhang}(2012)}]{wang12}%
  \BibitemOpen
  \bibfield  {author} {\bibinfo {author} {\bibfnamefont {Z.}~\bibnamefont
  {Wang}}\ and\ \bibinfo {author} {\bibfnamefont {S.-C.}\ \bibnamefont
  {Zhang}},\ }\href {\doibase 10.1103/PhysRevX.2.031008} {\bibfield  {journal}
  {\bibinfo  {journal} {Phys. Rev. X}\ }\textbf {\bibinfo {volume} {2}},\
  \bibinfo {pages} {031008} (\bibinfo {year} {2012})}\BibitemShut {NoStop}%
\bibitem [{\citenamefont {Wang}\ and\ \citenamefont {Yan}(2013)}]{Wang_2013}%
  \BibitemOpen
  \bibfield  {author} {\bibinfo {author} {\bibfnamefont {Z.}~\bibnamefont
  {Wang}}\ and\ \bibinfo {author} {\bibfnamefont {B.}~\bibnamefont {Yan}},\
  }\href {\doibase 10.1088/0953-8984/25/15/155601} {\bibfield  {journal}
  {\bibinfo  {journal} {Journal of Physics: Condensed Matter}\ }\textbf
  {\bibinfo {volume} {25}},\ \bibinfo {pages} {155601} (\bibinfo {year}
  {2013})}\BibitemShut {NoStop}%
\bibitem [{\citenamefont {Gros}\ and\ \citenamefont
  {Valent\'{\i}}(1993)}]{CPT-ValGros}%
  \BibitemOpen
  \bibfield  {author} {\bibinfo {author} {\bibfnamefont {C.}~\bibnamefont
  {Gros}}\ and\ \bibinfo {author} {\bibfnamefont {R.}~\bibnamefont
  {Valent\'{\i}}},\ }\href {\doibase 10.1103/PhysRevB.48.418} {\bibfield
  {journal} {\bibinfo  {journal} {Phys. Rev. B}\ }\textbf {\bibinfo {volume}
  {48}},\ \bibinfo {pages} {418} (\bibinfo {year} {1993})}\BibitemShut
  {NoStop}%
\bibitem [{\citenamefont {Gros}\ and\ \citenamefont
  {Valenti}(1994)}]{gros1994}%
  \BibitemOpen
  \bibfield  {author} {\bibinfo {author} {\bibfnamefont {C.}~\bibnamefont
  {Gros}}\ and\ \bibinfo {author} {\bibfnamefont {R.}~\bibnamefont {Valenti}},\
  }\href@noop {} {\bibfield  {journal} {\bibinfo  {journal} {Annalen der
  Physik}\ }\textbf {\bibinfo {volume} {506}},\ \bibinfo {pages} {460}
  (\bibinfo {year} {1994})}\BibitemShut {NoStop}%
\bibitem [{\citenamefont {S\'en\'echal}\ \emph
  {et~al.}(2000{\natexlab{a}})\citenamefont {S\'en\'echal}, \citenamefont
  {Perez},\ and\ \citenamefont {Pioro-Ladri\`ere}}]{PhysRevLett.84.522}%
  \BibitemOpen
  \bibfield  {author} {\bibinfo {author} {\bibfnamefont {D.}~\bibnamefont
  {S\'en\'echal}}, \bibinfo {author} {\bibfnamefont {D.}~\bibnamefont {Perez}},
  \ and\ \bibinfo {author} {\bibfnamefont {M.}~\bibnamefont
  {Pioro-Ladri\`ere}},\ }\href {\doibase 10.1103/PhysRevLett.84.522} {\bibfield
   {journal} {\bibinfo  {journal} {Phys. Rev. Lett.}\ }\textbf {\bibinfo
  {volume} {84}},\ \bibinfo {pages} {522} (\bibinfo {year}
  {2000}{\natexlab{a}})}\BibitemShut {NoStop}%
\bibitem [{\citenamefont {S\'en\'echal}\ \emph {et~al.}(2002)\citenamefont
  {S\'en\'echal}, \citenamefont {Perez},\ and\ \citenamefont
  {Plouffe}}]{CPT-Sene}%
  \BibitemOpen
  \bibfield  {author} {\bibinfo {author} {\bibfnamefont {D.}~\bibnamefont
  {S\'en\'echal}}, \bibinfo {author} {\bibfnamefont {D.}~\bibnamefont {Perez}},
  \ and\ \bibinfo {author} {\bibfnamefont {D.}~\bibnamefont {Plouffe}},\ }\href
  {\doibase 10.1103/PhysRevB.66.075129} {\bibfield  {journal} {\bibinfo
  {journal} {Phys. Rev. B}\ }\textbf {\bibinfo {volume} {66}},\ \bibinfo
  {pages} {075129} (\bibinfo {year} {2002})}\BibitemShut {NoStop}%
\bibitem [{\citenamefont {Manghi}(2013{\natexlab{a}})}]{Manghi_2013}%
  \BibitemOpen
  \bibfield  {author} {\bibinfo {author} {\bibfnamefont {F.}~\bibnamefont
  {Manghi}},\ }\href {\doibase 10.1088/0953-8984/26/1/015602} {\bibfield
  {journal} {\bibinfo  {journal} {Journal of Physics: Condensed Matter}\
  }\textbf {\bibinfo {volume} {26}},\ \bibinfo {pages} {015602} (\bibinfo
  {year} {2013}{\natexlab{a}})}\BibitemShut {NoStop}%
\bibitem [{\citenamefont {McCulloch}(2008{\natexlab{a}})}]{mcculloch2008}%
  \BibitemOpen
  \bibfield  {author} {\bibinfo {author} {\bibfnamefont {I.~P.}\ \bibnamefont
  {McCulloch}},\ }\href@noop {} {\bibfield  {journal} {\bibinfo  {journal}
  {arXiv preprint arXiv:0804.2509}\ } (\bibinfo {year}
  {2008}{\natexlab{a}})}\BibitemShut {NoStop}%
\bibitem [{\citenamefont {Zaletel}\ \emph {et~al.}(2013)\citenamefont
  {Zaletel}, \citenamefont {Mong},\ and\ \citenamefont
  {Pollmann}}]{zaletel2013}%
  \BibitemOpen
  \bibfield  {author} {\bibinfo {author} {\bibfnamefont {M.~P.}\ \bibnamefont
  {Zaletel}}, \bibinfo {author} {\bibfnamefont {R.~S.}\ \bibnamefont {Mong}}, \
  and\ \bibinfo {author} {\bibfnamefont {F.}~\bibnamefont {Pollmann}},\
  }\href@noop {} {\bibfield  {journal} {\bibinfo  {journal} {Physical review
  letters}\ }\textbf {\bibinfo {volume} {110}},\ \bibinfo {pages} {236801}
  (\bibinfo {year} {2013})}\BibitemShut {NoStop}%
\bibitem [{\citenamefont {Becca}\ and\ \citenamefont
  {Sorella}(2017)}]{becca2017}%
  \BibitemOpen
  \bibfield  {author} {\bibinfo {author} {\bibfnamefont {F.}~\bibnamefont
  {Becca}}\ and\ \bibinfo {author} {\bibfnamefont {S.}~\bibnamefont
  {Sorella}},\ }\href {\doibase 10.1017/9781316417041} {\emph {\bibinfo {title}
  {Quantum Monte Carlo Approaches for Correlated Systems}}}\ (\bibinfo
  {publisher} {Cambridge University Press},\ \bibinfo {year}
  {2017})\BibitemShut {NoStop}%
\bibitem [{\citenamefont {Capello}\ \emph {et~al.}(2005)\citenamefont
  {Capello}, \citenamefont {Becca}, \citenamefont {Fabrizio}, \citenamefont
  {Sorella},\ and\ \citenamefont {Tosatti}}]{capello2005}%
  \BibitemOpen
  \bibfield  {author} {\bibinfo {author} {\bibfnamefont {M.}~\bibnamefont
  {Capello}}, \bibinfo {author} {\bibfnamefont {F.}~\bibnamefont {Becca}},
  \bibinfo {author} {\bibfnamefont {M.}~\bibnamefont {Fabrizio}}, \bibinfo
  {author} {\bibfnamefont {S.}~\bibnamefont {Sorella}}, \ and\ \bibinfo
  {author} {\bibfnamefont {E.}~\bibnamefont {Tosatti}},\ }\href {\doibase
  10.1103/PhysRevLett.94.026406} {\bibfield  {journal} {\bibinfo  {journal}
  {Phys. Rev. Lett.}\ }\textbf {\bibinfo {volume} {94}},\ \bibinfo {pages}
  {026406} (\bibinfo {year} {2005})}\BibitemShut {NoStop}%
\bibitem [{\citenamefont {Capello}\ \emph {et~al.}(2006)\citenamefont
  {Capello}, \citenamefont {Becca}, \citenamefont {Yunoki},\ and\ \citenamefont
  {Sorella}}]{capello2006}%
  \BibitemOpen
  \bibfield  {author} {\bibinfo {author} {\bibfnamefont {M.}~\bibnamefont
  {Capello}}, \bibinfo {author} {\bibfnamefont {F.}~\bibnamefont {Becca}},
  \bibinfo {author} {\bibfnamefont {S.}~\bibnamefont {Yunoki}}, \ and\ \bibinfo
  {author} {\bibfnamefont {S.}~\bibnamefont {Sorella}},\ }\href {\doibase
  10.1103/PhysRevB.73.245116} {\bibfield  {journal} {\bibinfo  {journal} {Phys.
  Rev. B}\ }\textbf {\bibinfo {volume} {73}},\ \bibinfo {pages} {245116}
  (\bibinfo {year} {2006})}\BibitemShut {NoStop}%
\bibitem [{\citenamefont {Tocchio}\ \emph {et~al.}(2014)\citenamefont
  {Tocchio}, \citenamefont {Gros}, \citenamefont {Valent{\'\i}},\ and\
  \citenamefont {Becca}}]{tocchio2014}%
  \BibitemOpen
  \bibfield  {author} {\bibinfo {author} {\bibfnamefont {L.~F.}\ \bibnamefont
  {Tocchio}}, \bibinfo {author} {\bibfnamefont {C.}~\bibnamefont {Gros}},
  \bibinfo {author} {\bibfnamefont {R.}~\bibnamefont {Valent{\'\i}}}, \ and\
  \bibinfo {author} {\bibfnamefont {F.}~\bibnamefont {Becca}},\ }\href@noop {}
  {\bibfield  {journal} {\bibinfo  {journal} {Physical Review B}\ }\textbf
  {\bibinfo {volume} {89}},\ \bibinfo {pages} {235107} (\bibinfo {year}
  {2014})}\BibitemShut {NoStop}%
\bibitem [{\citenamefont {Pollmann}\ and\ \citenamefont
  {Turner}(2012{\natexlab{a}})}]{Pollmann_2012}%
  \BibitemOpen
  \bibfield  {author} {\bibinfo {author} {\bibfnamefont {F.}~\bibnamefont
  {Pollmann}}\ and\ \bibinfo {author} {\bibfnamefont {A.~M.}\ \bibnamefont
  {Turner}},\ }\href {\doibase 10.1103/PhysRevB.86.125441} {\bibfield
  {journal} {\bibinfo  {journal} {Phys. Rev. B}\ }\textbf {\bibinfo {volume}
  {86}},\ \bibinfo {pages} {125441} (\bibinfo {year}
  {2012}{\natexlab{a}})}\BibitemShut {NoStop}%
\bibitem [{\citenamefont {Aroyo}\ \emph {et~al.}(2006)\citenamefont {Aroyo},
  \citenamefont {Perez-Mato}, \citenamefont {Capillas}, \citenamefont
  {Kroumova}, \citenamefont {Ivantchev}, \citenamefont {Madariaga},
  \citenamefont {Kirov},\ and\ \citenamefont {Wondratschek}}]{BCS1}%
  \BibitemOpen
  \bibfield  {author} {\bibinfo {author} {\bibfnamefont {M.~I.}\ \bibnamefont
  {Aroyo}}, \bibinfo {author} {\bibfnamefont {J.~M.}\ \bibnamefont
  {Perez-Mato}}, \bibinfo {author} {\bibfnamefont {C.}~\bibnamefont
  {Capillas}}, \bibinfo {author} {\bibfnamefont {E.}~\bibnamefont {Kroumova}},
  \bibinfo {author} {\bibfnamefont {S.}~\bibnamefont {Ivantchev}}, \bibinfo
  {author} {\bibfnamefont {G.}~\bibnamefont {Madariaga}}, \bibinfo {author}
  {\bibfnamefont {A.}~\bibnamefont {Kirov}}, \ and\ \bibinfo {author}
  {\bibfnamefont {H.}~\bibnamefont {Wondratschek}},\ }\href
  {https://www.degruyter.com/view/journals/zkri/221/1/article-p15.xml}
  {\bibfield  {journal} {\bibinfo  {journal} {Zeitschrift f{\"u}r
  Kristallographie - Crystalline Materials}\ }\textbf {\bibinfo {volume}
  {221}},\ \bibinfo {pages} {15 } (\bibinfo {year} {2006})}\BibitemShut
  {NoStop}%
\bibitem [{\citenamefont {Aroyo}\ \emph {et~al.}(2011)\citenamefont {Aroyo},
  \citenamefont {Perez-Mato}, \citenamefont {Orobengoa}, \citenamefont {Tasci},
  \citenamefont {de~la Flor},\ and\ \citenamefont {Kirov}}]{BCS2}%
  \BibitemOpen
  \bibfield  {author} {\bibinfo {author} {\bibfnamefont {M.~I.}\ \bibnamefont
  {Aroyo}}, \bibinfo {author} {\bibfnamefont {J.~M.}\ \bibnamefont
  {Perez-Mato}}, \bibinfo {author} {\bibfnamefont {D.}~\bibnamefont
  {Orobengoa}}, \bibinfo {author} {\bibfnamefont {E.}~\bibnamefont {Tasci}},
  \bibinfo {author} {\bibfnamefont {G.}~\bibnamefont {de~la Flor}}, \ and\
  \bibinfo {author} {\bibfnamefont {A.}~\bibnamefont {Kirov}},\ }\href@noop {}
  {\bibfield  {journal} {\bibinfo  {journal} {Bulg. Chem. Commun.}\ }\textbf
  {\bibinfo {volume} {43}},\ \bibinfo {pages} {183} (\bibinfo {year}
  {2011})}\BibitemShut {NoStop}%
\bibitem [{C2v()}]{C2v_vs_D2h}%
  \BibitemOpen
  \href@noop {} {}\bibinfo {note} {Note that C$_{2v}$ is a subgroup of
  D$_{2h}$, where D$_{2h}$ is the point group of an isolated
  diamond.}\BibitemShut {Stop}%
\bibitem [{pos()}]{pos_hop}%
  \BibitemOpen
  \href@noop {} {}\bibinfo {note} {Although the choice of restricting our
  analysis to positive hopping parameters means that we can not study all
  insulator phases of the model, it gives us access to the part of the phase
  diagram containing the physics we are interested in, namely two different
  topological phases distinguished by symmetry and a Mott phase.}\BibitemShut
  {Stop}%
\bibitem [{low()}]{lowestband}%
  \BibitemOpen
  \href@noop {} {}\bibinfo {note} {Based on the similarity of the correlated
  insulator phases and the noninteracting AI and OAL phases, we may predict the
  possibility of the crossing between the lowest and last occupied bands of the
  topological Hamiltonian in the mentioned phases. Such a interplay of bands
  would not have any effect in the topological classification.}\BibitemShut
  {Stop}%
\bibitem [{\citenamefont {Hastings}\ and\ \citenamefont
  {Koma}(2006)}]{hastings2006spectral}%
  \BibitemOpen
  \bibfield  {author} {\bibinfo {author} {\bibfnamefont {M.~B.}\ \bibnamefont
  {Hastings}}\ and\ \bibinfo {author} {\bibfnamefont {T.}~\bibnamefont
  {Koma}},\ }\href@noop {} {\bibfield  {journal} {\bibinfo  {journal}
  {Communications in mathematical physics}\ }\textbf {\bibinfo {volume}
  {265}},\ \bibinfo {pages} {781} (\bibinfo {year} {2006})}\BibitemShut
  {NoStop}%
\bibitem [{\citenamefont {Pollmann}\ and\ \citenamefont
  {Turner}(2012{\natexlab{b}})}]{pollmann2012det}%
  \BibitemOpen
  \bibfield  {author} {\bibinfo {author} {\bibfnamefont {F.}~\bibnamefont
  {Pollmann}}\ and\ \bibinfo {author} {\bibfnamefont {A.~M.}\ \bibnamefont
  {Turner}},\ }\href@noop {} {\bibfield  {journal} {\bibinfo  {journal}
  {Physical review b}\ }\textbf {\bibinfo {volume} {86}},\ \bibinfo {pages}
  {125441} (\bibinfo {year} {2012}{\natexlab{b}})}\BibitemShut {NoStop}%
\bibitem [{\citenamefont {Shiozaki}\ \emph {et~al.}(2017)\citenamefont
  {Shiozaki}, \citenamefont {Shapourian},\ and\ \citenamefont
  {Ryu}}]{Shiozaki_2017}%
  \BibitemOpen
  \bibfield  {author} {\bibinfo {author} {\bibfnamefont {K.}~\bibnamefont
  {Shiozaki}}, \bibinfo {author} {\bibfnamefont {H.}~\bibnamefont
  {Shapourian}}, \ and\ \bibinfo {author} {\bibfnamefont {S.}~\bibnamefont
  {Ryu}},\ }\href {\doibase 10.1103/PhysRevB.95.205139} {\bibfield  {journal}
  {\bibinfo  {journal} {Phys. Rev. B}\ }\textbf {\bibinfo {volume} {95}},\
  \bibinfo {pages} {205139} (\bibinfo {year} {2017})}\BibitemShut {NoStop}%
\bibitem [{\citenamefont {Shapourian}\ \emph {et~al.}(2017)\citenamefont
  {Shapourian}, \citenamefont {Shiozaki},\ and\ \citenamefont
  {Ryu}}]{Shiozaki_Manybody_2017}%
  \BibitemOpen
  \bibfield  {author} {\bibinfo {author} {\bibfnamefont {H.}~\bibnamefont
  {Shapourian}}, \bibinfo {author} {\bibfnamefont {K.}~\bibnamefont
  {Shiozaki}}, \ and\ \bibinfo {author} {\bibfnamefont {S.}~\bibnamefont
  {Ryu}},\ }\href {\doibase 10.1103/PhysRevLett.118.216402} {\bibfield
  {journal} {\bibinfo  {journal} {Phys. Rev. Lett.}\ }\textbf {\bibinfo
  {volume} {118}},\ \bibinfo {pages} {216402} (\bibinfo {year}
  {2017})}\BibitemShut {NoStop}%
\bibitem [{\citenamefont {Volovik}(2003)}]{Volovik2003}%
  \BibitemOpen
  \bibfield  {author} {\bibinfo {author} {\bibfnamefont {G.}~\bibnamefont
  {Volovik}},\ }\href@noop {} {\emph {\bibinfo {title} {The Universe in a
  helium droplet}}}\ (\bibinfo  {publisher} {Oxford University Press},\
  \bibinfo {year} {2003})\BibitemShut {NoStop}%
\bibitem [{\citenamefont {Wang}\ \emph {et~al.}(2010)\citenamefont {Wang},
  \citenamefont {Qi},\ and\ \citenamefont {Zhang}}]{wang10}%
  \BibitemOpen
  \bibfield  {author} {\bibinfo {author} {\bibfnamefont {Z.}~\bibnamefont
  {Wang}}, \bibinfo {author} {\bibfnamefont {X.-L.}\ \bibnamefont {Qi}}, \ and\
  \bibinfo {author} {\bibfnamefont {S.-C.}\ \bibnamefont {Zhang}},\ }\href
  {\doibase 10.1103/PhysRevLett.105.256803} {\bibfield  {journal} {\bibinfo
  {journal} {Phys. Rev. Lett.}\ }\textbf {\bibinfo {volume} {105}},\ \bibinfo
  {pages} {256803} (\bibinfo {year} {2010})}\BibitemShut {NoStop}%
\bibitem [{\citenamefont {Wang}\ \emph {et~al.}(2012)\citenamefont {Wang},
  \citenamefont {Qi},\ and\ \citenamefont {Zhang}}]{wang12inv}%
  \BibitemOpen
  \bibfield  {author} {\bibinfo {author} {\bibfnamefont {Z.}~\bibnamefont
  {Wang}}, \bibinfo {author} {\bibfnamefont {X.-L.}\ \bibnamefont {Qi}}, \ and\
  \bibinfo {author} {\bibfnamefont {S.-C.}\ \bibnamefont {Zhang}},\ }\href
  {\doibase 10.1103/PhysRevB.85.165126} {\bibfield  {journal} {\bibinfo
  {journal} {Phys. Rev. B}\ }\textbf {\bibinfo {volume} {85}},\ \bibinfo
  {pages} {165126} (\bibinfo {year} {2012})}\BibitemShut {NoStop}%
\bibitem [{\citenamefont {Mertz}\ \emph {et~al.}(2019)\citenamefont {Mertz},
  \citenamefont {Zantout},\ and\ \citenamefont {Valent{\'\i}}}]{mertz2019}%
  \BibitemOpen
  \bibfield  {author} {\bibinfo {author} {\bibfnamefont {T.}~\bibnamefont
  {Mertz}}, \bibinfo {author} {\bibfnamefont {K.}~\bibnamefont {Zantout}}, \
  and\ \bibinfo {author} {\bibfnamefont {R.}~\bibnamefont {Valent{\'\i}}},\
  }\href@noop {} {\bibfield  {journal} {\bibinfo  {journal} {Physical Review
  B}\ }\textbf {\bibinfo {volume} {100}},\ \bibinfo {pages} {125111} (\bibinfo
  {year} {2019})}\BibitemShut {NoStop}%
\bibitem [{\citenamefont {Lessnich}()}]{Lessnich2021}%
  \BibitemOpen
  \bibfield  {author} {\bibinfo {author} {\bibfnamefont {D.}~\bibnamefont
  {Lessnich}},\ }\href@noop {} {}\bibinfo {howpublished} {\textit{et al. in
  preparation}}\BibitemShut {NoStop}%
\bibitem [{\citenamefont {Manmana}\ \emph {et~al.}(2012)\citenamefont
  {Manmana}, \citenamefont {Essin}, \citenamefont {Noack},\ and\ \citenamefont
  {Gurarie}}]{gurarie1d}%
  \BibitemOpen
  \bibfield  {author} {\bibinfo {author} {\bibfnamefont {S.~R.}\ \bibnamefont
  {Manmana}}, \bibinfo {author} {\bibfnamefont {A.~M.}\ \bibnamefont {Essin}},
  \bibinfo {author} {\bibfnamefont {R.~M.}\ \bibnamefont {Noack}}, \ and\
  \bibinfo {author} {\bibfnamefont {V.}~\bibnamefont {Gurarie}},\ }\href
  {\doibase 10.1103/PhysRevB.86.205119} {\bibfield  {journal} {\bibinfo
  {journal} {Phys. Rev. B}\ }\textbf {\bibinfo {volume} {86}},\ \bibinfo
  {pages} {205119} (\bibinfo {year} {2012})}\BibitemShut {NoStop}%
\bibitem [{\citenamefont {He}\ \emph {et~al.}(2016)\citenamefont {He},
  \citenamefont {Wu}, \citenamefont {Meng},\ and\ \citenamefont
  {Lu}}]{limitG2}%
  \BibitemOpen
  \bibfield  {author} {\bibinfo {author} {\bibfnamefont {Y.-Y.}\ \bibnamefont
  {He}}, \bibinfo {author} {\bibfnamefont {H.-Q.}\ \bibnamefont {Wu}}, \bibinfo
  {author} {\bibfnamefont {Z.~Y.}\ \bibnamefont {Meng}}, \ and\ \bibinfo
  {author} {\bibfnamefont {Z.-Y.}\ \bibnamefont {Lu}},\ }\href {\doibase
  10.1103/PhysRevB.93.195164} {\bibfield  {journal} {\bibinfo  {journal} {Phys.
  Rev. B}\ }\textbf {\bibinfo {volume} {93}},\ \bibinfo {pages} {195164}
  (\bibinfo {year} {2016})}\BibitemShut {NoStop}%
\bibitem [{\citenamefont {You}\ \emph {et~al.}(2014)\citenamefont {You},
  \citenamefont {Wang}, \citenamefont {Oon},\ and\ \citenamefont
  {Xu}}]{unkonv_pert}%
  \BibitemOpen
  \bibfield  {author} {\bibinfo {author} {\bibfnamefont {Y.-Z.}\ \bibnamefont
  {You}}, \bibinfo {author} {\bibfnamefont {Z.}~\bibnamefont {Wang}}, \bibinfo
  {author} {\bibfnamefont {J.}~\bibnamefont {Oon}}, \ and\ \bibinfo {author}
  {\bibfnamefont {C.}~\bibnamefont {Xu}},\ }\href {\doibase
  10.1103/PhysRevB.90.060502} {\bibfield  {journal} {\bibinfo  {journal} {Phys.
  Rev. B}\ }\textbf {\bibinfo {volume} {90}},\ \bibinfo {pages} {060502}
  (\bibinfo {year} {2014})}\BibitemShut {NoStop}%
\bibitem [{\citenamefont {White}(1992)}]{white1992density}%
  \BibitemOpen
  \bibfield  {author} {\bibinfo {author} {\bibfnamefont {S.~R.}\ \bibnamefont
  {White}},\ }\href@noop {} {\bibfield  {journal} {\bibinfo  {journal}
  {Physical review letters}\ }\textbf {\bibinfo {volume} {69}},\ \bibinfo
  {pages} {2863} (\bibinfo {year} {1992})}\BibitemShut {NoStop}%
\bibitem [{\citenamefont {White}(1993)}]{white1993density}%
  \BibitemOpen
  \bibfield  {author} {\bibinfo {author} {\bibfnamefont {S.~R.}\ \bibnamefont
  {White}},\ }\href@noop {} {\bibfield  {journal} {\bibinfo  {journal}
  {Physical Review B}\ }\textbf {\bibinfo {volume} {48}},\ \bibinfo {pages}
  {10345} (\bibinfo {year} {1993})}\BibitemShut {NoStop}%
\bibitem [{\citenamefont
  {McCulloch}(2008{\natexlab{b}})}]{mcculloch2008infinite}%
  \BibitemOpen
  \bibfield  {author} {\bibinfo {author} {\bibfnamefont {I.}~\bibnamefont
  {McCulloch}},\ }\href@noop {} {\bibfield  {journal} {\bibinfo  {journal}
  {arXiv preprint arXiv:0804.2509}\ } (\bibinfo {year}
  {2008}{\natexlab{b}})}\BibitemShut {NoStop}%
\bibitem [{\citenamefont {Hauschild}\ and\ \citenamefont
  {Pollmann}(2018)}]{hauschild2018efficient}%
  \BibitemOpen
  \bibfield  {author} {\bibinfo {author} {\bibfnamefont {J.}~\bibnamefont
  {Hauschild}}\ and\ \bibinfo {author} {\bibfnamefont {F.}~\bibnamefont
  {Pollmann}},\ }\href@noop {} {\bibfield  {journal} {\bibinfo  {journal}
  {SciPost Physics Lecture Notes}\ } (\bibinfo {year} {2018})}\BibitemShut
  {NoStop}%
\bibitem [{\citenamefont {Sorella}(2005)}]{sorella2005}%
  \BibitemOpen
  \bibfield  {author} {\bibinfo {author} {\bibfnamefont {S.}~\bibnamefont
  {Sorella}},\ }\href {\doibase 10.1103/PhysRevB.71.241103} {\bibfield
  {journal} {\bibinfo  {journal} {Phys. Rev. B}\ }\textbf {\bibinfo {volume}
  {71}},\ \bibinfo {pages} {241103} (\bibinfo {year} {2005})}\BibitemShut
  {NoStop}%
\bibitem [{\citenamefont {Feynman}(1954)}]{feynman1954}%
  \BibitemOpen
  \bibfield  {author} {\bibinfo {author} {\bibfnamefont {R.~P.}\ \bibnamefont
  {Feynman}},\ }\href {\doibase 10.1103/PhysRev.94.262} {\bibfield  {journal}
  {\bibinfo  {journal} {Phys. Rev.}\ }\textbf {\bibinfo {volume} {94}},\
  \bibinfo {pages} {262} (\bibinfo {year} {1954})}\BibitemShut {NoStop}%
\bibitem [{\citenamefont {Resta}\ and\ \citenamefont
  {Sorella}(1999)}]{resta1999}%
  \BibitemOpen
  \bibfield  {author} {\bibinfo {author} {\bibfnamefont {R.}~\bibnamefont
  {Resta}}\ and\ \bibinfo {author} {\bibfnamefont {S.}~\bibnamefont
  {Sorella}},\ }\href {\doibase 10.1103/PhysRevLett.82.370} {\bibfield
  {journal} {\bibinfo  {journal} {Phys. Rev. Lett.}\ }\textbf {\bibinfo
  {volume} {82}},\ \bibinfo {pages} {370} (\bibinfo {year} {1999})}\BibitemShut
  {NoStop}%
\bibitem [{\citenamefont {S\'en\'echal}\ \emph
  {et~al.}(2000{\natexlab{b}})\citenamefont {S\'en\'echal}, \citenamefont
  {Perez},\ and\ \citenamefont {Pioro-Ladri\`ere}}]{CPT-Sene1}%
  \BibitemOpen
  \bibfield  {author} {\bibinfo {author} {\bibfnamefont {D.}~\bibnamefont
  {S\'en\'echal}}, \bibinfo {author} {\bibfnamefont {D.}~\bibnamefont {Perez}},
  \ and\ \bibinfo {author} {\bibfnamefont {M.}~\bibnamefont
  {Pioro-Ladri\`ere}},\ }\href {\doibase 10.1103/PhysRevLett.84.522} {\bibfield
   {journal} {\bibinfo  {journal} {Phys. Rev. Lett.}\ }\textbf {\bibinfo
  {volume} {84}},\ \bibinfo {pages} {522} (\bibinfo {year}
  {2000}{\natexlab{b}})}\BibitemShut {NoStop}%
\bibitem [{\citenamefont {Sénéchal}(2010)}]{CPT-Sene2}%
  \BibitemOpen
  \bibfield  {author} {\bibinfo {author} {\bibfnamefont {D.}~\bibnamefont
  {Sénéchal}},\ }\href@noop {} {\enquote {\bibinfo {title} {An introduction
  to quantum cluster methods},}\ } (\bibinfo {year} {2010}),\ \Eprint
  {http://arxiv.org/abs/0806.2690} {arXiv:0806.2690 [cond-mat.str-el]}
  \BibitemShut {NoStop}%
\bibitem [{\citenamefont {Manghi}(2013{\natexlab{b}})}]{CPT-Manghi}%
  \BibitemOpen
  \bibfield  {author} {\bibinfo {author} {\bibfnamefont {F.}~\bibnamefont
  {Manghi}},\ }\href {\doibase 10.1088/0953-8984/26/1/015602} {\bibfield
  {journal} {\bibinfo  {journal} {Journal of Physics: Condensed Matter}\
  }\textbf {\bibinfo {volume} {26}},\ \bibinfo {pages} {015602} (\bibinfo
  {year} {2013}{\natexlab{b}})}\BibitemShut {NoStop}%
\bibitem [{\citenamefont {Dagotto}(1994)}]{ED1}%
  \BibitemOpen
  \bibfield  {author} {\bibinfo {author} {\bibfnamefont {E.}~\bibnamefont
  {Dagotto}},\ }\href {\doibase 10.1103/RevModPhys.66.763} {\bibfield
  {journal} {\bibinfo  {journal} {Rev. Mod. Phys.}\ }\textbf {\bibinfo {volume}
  {66}},\ \bibinfo {pages} {763} (\bibinfo {year} {1994})}\BibitemShut
  {NoStop}%
\bibitem [{\citenamefont {Pairault}\ \emph {et~al.}(1998)\citenamefont
  {Pairault}, \citenamefont {S\'en\'echal},\ and\ \citenamefont
  {Tremblay}}]{CPT-Tremblay}%
  \BibitemOpen
  \bibfield  {author} {\bibinfo {author} {\bibfnamefont {S.}~\bibnamefont
  {Pairault}}, \bibinfo {author} {\bibfnamefont {D.}~\bibnamefont
  {S\'en\'echal}}, \ and\ \bibinfo {author} {\bibfnamefont {A.-M.~S.}\
  \bibnamefont {Tremblay}},\ }\href {\doibase 10.1103/PhysRevLett.80.5389}
  {\bibfield  {journal} {\bibinfo  {journal} {Phys. Rev. Lett.}\ }\textbf
  {\bibinfo {volume} {80}},\ \bibinfo {pages} {5389} (\bibinfo {year}
  {1998})}\BibitemShut {NoStop}%
\end{thebibliography}%

\end{document}